\documentclass{aa}
\usepackage{graphicx}
\usepackage[varg]{txfonts}
\usepackage{amssymb}            
\usepackage{color}              
\newcommand{\etal}{et al. }
\newcommand{\be}{\begin{equation}}
\newcommand{\ee}{\end{equation}}
\newcommand{\beq}{\begin{eqnarray}}
\newcommand{\eeq}{\end{eqnarray}}
\newcommand{\RSUN}{$R_{\odot}$}

\renewcommand{\bf}{}

\begin{document}

 \title{High resolution observations with Artemis-IV and the NRH}
 \subtitle{I. Type IV associated  narrow-band bursts}

   \author{                     C. Bouratzis \inst{1}
                                                        A. Hillaris \inst{1}
                                                        C.E. Alissandrakis \inst{2}
                                                        P. Preka-Papadema \inst{1}
                                                        X. Moussas \inst{1}
                                                        C. Caroubalos \inst{3}\\
                                                        P. Tsitsipis \inst{4},
   \and
                                                        A. Kontogeorgos \inst{4} }
\offprints{C. Bouratzis}
\institute{                     Department of Physics, University of Athens, 15783 Athens, Greece\\
\email{kosbour@gmail.com}
\and                            Department of Physics, University of Ioannina, 45110 Ioannina, Greece
\and                            Department of Informatics, University of Athens, 15783 Athens, Greece
\and                                    Department of Electronics, Technological Educational Institute of Lamia, 35100 Lamia, Greece}
\authorrunning{Bouratzis et al.}
 \date{Received .....; accepted ......}
  \abstract
{Narrow-band bursts appear on dynamic spectra from microwave to decametric frequencies as fine structures with very small duration and bandwidth. They are believed to be manifestations of small scale energy release through magnetic reconnection.}    
{We analyzed 27 metric type IV events with embedded narrow-band bursts, which was observed by the ARTEMIS--IV radio spectrograph from 30 June 1999 to 1 August 2010. We examined the morphological characteristics of isolated narrow-band structures (mostly spikes) and groups or chains of structures.} 
{The events were recorded with the SAO high resolution (10 ms cadence) receiver of ARTEMIS-IV in the 270--450 MHz range. We measured the duration, spectral width, and frequency drift of $\sim$12\,000 individual narrow-band bursts, groups, and chains. Spike sources were imaged with the Nan\c cay radioheliograph (NRH) for the event of 21 April 2003.}
{\bf {The mean duration of individual bursts at fixed frequency was $\sim100$~ms, while the instantaneous relative bandwidth was $\sim2\%$. Some bursts had measurable frequency drift, either positive or negative. Quite often spikes appeared in chains, which were closely spaced in time (column chains) or in frequency (row chains). Column chains had frequency drifts similar to type-IIId bursts, while most of the row chains exhibited negative frequently drifts with a rate close to that of fiber bursts. From the analysis of NRH data, we found that spikes were superimposed on a larger, slowly varying, background component. They were polarized in the same sense as the background source, with a slightly higher degree of polarization of $\sim65$\%, and their size was about 60\% of their size in total intensity.}}
{\bf{The duration and bandwidth distributions did not show any clear separation in groups. Some chains tended to assume the form of zebra, lace stripes, fiber bursts, or bursts of the type-III family, suggesting that such bursts might be resolved in spikes when viewed with high resolution. The NRH data indicate that the spikes are not fluctuations of the background, but represent additional emission such as what would be expected from small-scale reconnection.} }
  
  \keywords{Sun: corona -- Sun: radio radiation -- Sun: Magnetic reconnection -- Sun: Acceleration of particles}
   \maketitle

%
\begin{figure}
\begin{center}
\includegraphics[trim=1.25cm 0.5cm 0.1cm 0cm,clip,keepaspectratio=true,width=\hsize]{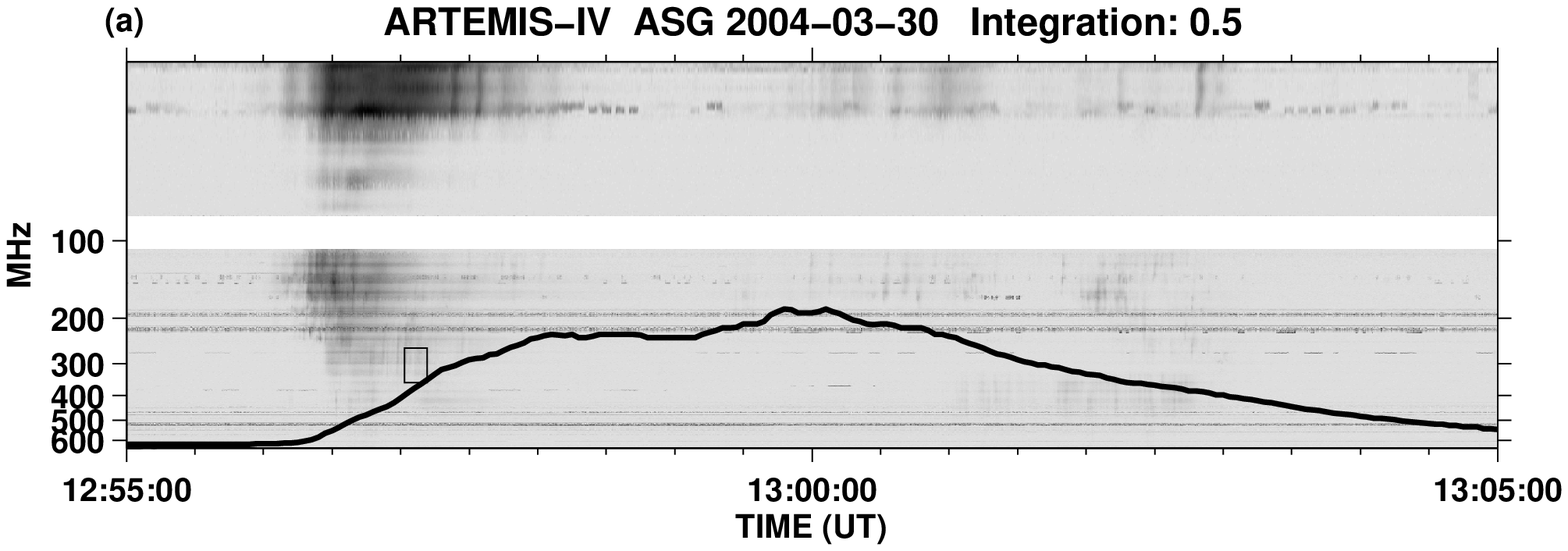}\\
\includegraphics[trim=0.0cm 0.5cm 0.1cm 0cm,clip,keepaspectratio=true,width=\hsize]{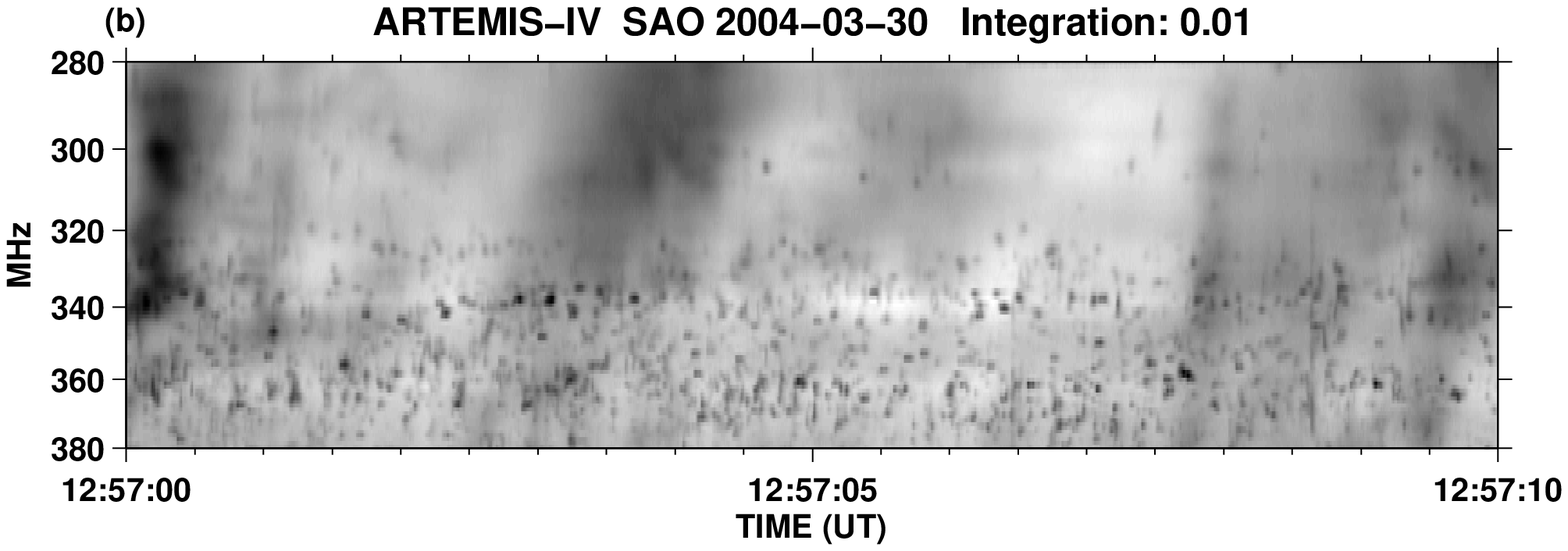}
\caption{{Typical example of narrow-band structures embedded in a complex event. (a): Low time resolution sweep-frequency spectrum from the ARTEMIS-IV/ASG receiver. (b): Detailed dynamic spectrum of the marked area in panel (a) recorded with the SAO with time resolution (10\.ms) receiver.}}
\label{Overall01}
\end{center}
\end{figure}

\section{Introduction}
Narrow-band bursts are a general class of fine structures including, but not restricted to, the narrow--band members of the type-III family, spikes, dots and sub-second patches, depending on their shape on the dynamic spectra; part of the same family are the III(U) and III(J) narrow--band bursts reported by \citet{Fu04} and \citet{Bouratzis10}. They have been interpreted as signatures of small scale acceleration episodes \citep[e.g.,][]{Nindos07}. In this basic class we might also include the \emph{Sawtooth oscillations} by \cite{Klassen01}, associated with type--II shocks. 

Spike bursts \citep{Benz1986} represent a very large part of the short duration ($\delta \tau$ of  tens of ms), narrow-band ($\delta \ln f~\approx~1-2\%$ in the 0.5--15 MHz range) radio bursts. The typical duration of a single spike decreases with observing frequency: the decametric spikes were found to have durations of 1200 ms at 18--30 MHz \citep{Melnik2011}, while in higher frequencies ranges their duration decreases from 50--100 ms at 250 MHz to 10--50 ms at 460 MHz, reaching 3--7 ms at 1420 MHz. 

The adjacency in time and frequency of  the  short and narrow-band  spikes to the starting time and frequency of type-III bursts groups was noted by \citet{Benz1982, Benz1996}. The latter proposed that this association was consistent with energy 
 release at high altitudes; the resulting suprathermal electron populations might provide the exciters of both the type IIIs and the 
 spikes.
In addition to type III, association of spikes with type-I and type-IV continua, as well as microwave bursts has been reported \citep[see, e.g.,][]{Benz1986,Chernov2011}.

Very often spikes exhibit measurable frequency drift rate. Spikes drifting toward lower frequencies (negative drift rate) are considered signatures of an outwardly moving exciter and are called \emph{direct} (d). Spikes with positive drift are named \emph{reverse} (r), while those without measurable drift rate are called \emph{stable} (s) \citep[see][]{Chernov2011}.
Spikes may appearing isolated, in loose clusters, or organized chains. They exhibit high brightness temperature, $T_B\sim 10 ^ { 15}$\,K in decimetric wavelengths \citep{Benz1986}. 

In a recent work \citep{Bouratzis2015}, we  found a very good temporal association between energy release episodes and spikes. We used the time of first impulsive energy release, shown by  the first HXR/microwave peak,  as  the reference for timing the appearance of fine structures with respect to the evolution of the flare process. Narrow-band bursts tend to cluster around the first impulsive energy release. Their histogram of time delay had a peak at 0.0\,min with a median value of 6.5\,min and a full width at half maximum (FWHM) of 18.0\,min (see their Figure 7). Although they concentrate mostly around the flare maximum, spikes occasionally cover longer periods during the flare decay phase, and are probably associated with subsequent energy releases. Narrow-band fine structures may also appear before the impulsive phase of the flare \citep{Aurass2007}.

In this work we examine narrow-band bursts and spikes observed during type-IV events recorded by the Artemis IV solar radio-spectrograph from June 30, 1999 until August 1, 2010 in the 270-450 MHz frequency range.  We study the morphological characteristics of isolated members of this type of fine structure, as well as of groups and chains. In Sect. \ref{Obs} we present the observations and their analysis, in Sect. \ref{Res} we discuss the results, in Sect. \ref{NRH} we analyze two-dimensional observations with the Nan\c cay Radioheliograph,  and in Sect. \ref{discussion} we present  our conclusions.

\begin{figure}[h]
\begin{center}
\includegraphics[width=\hsize]{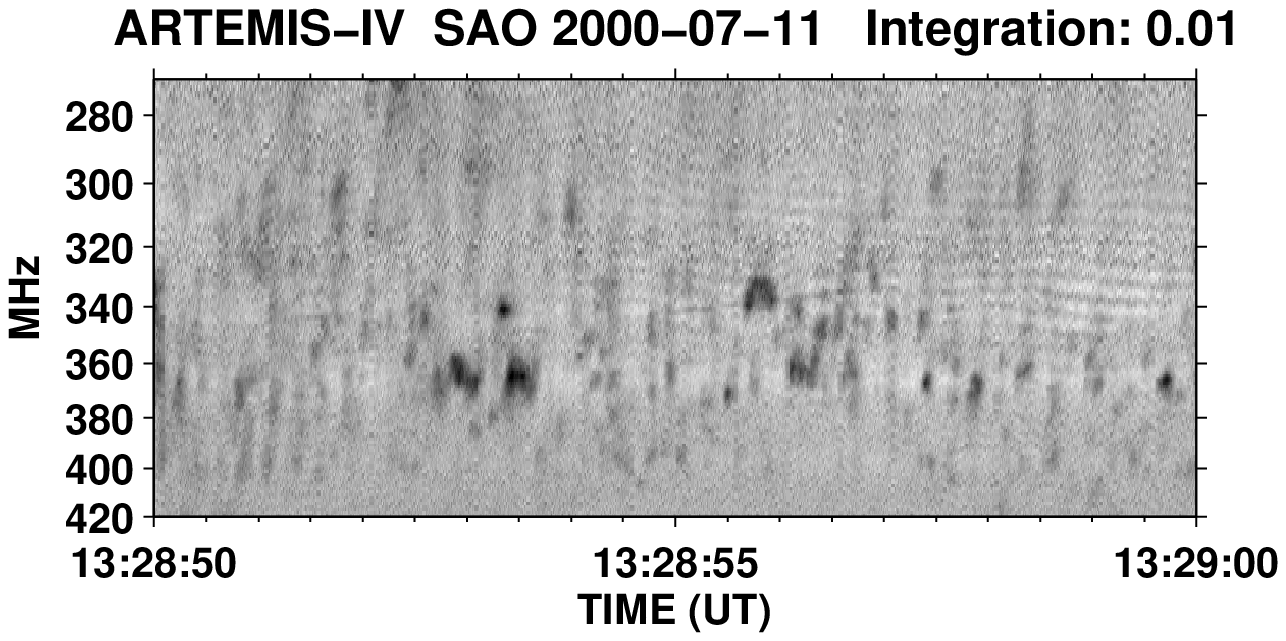} \\ 
\includegraphics[width=\hsize]{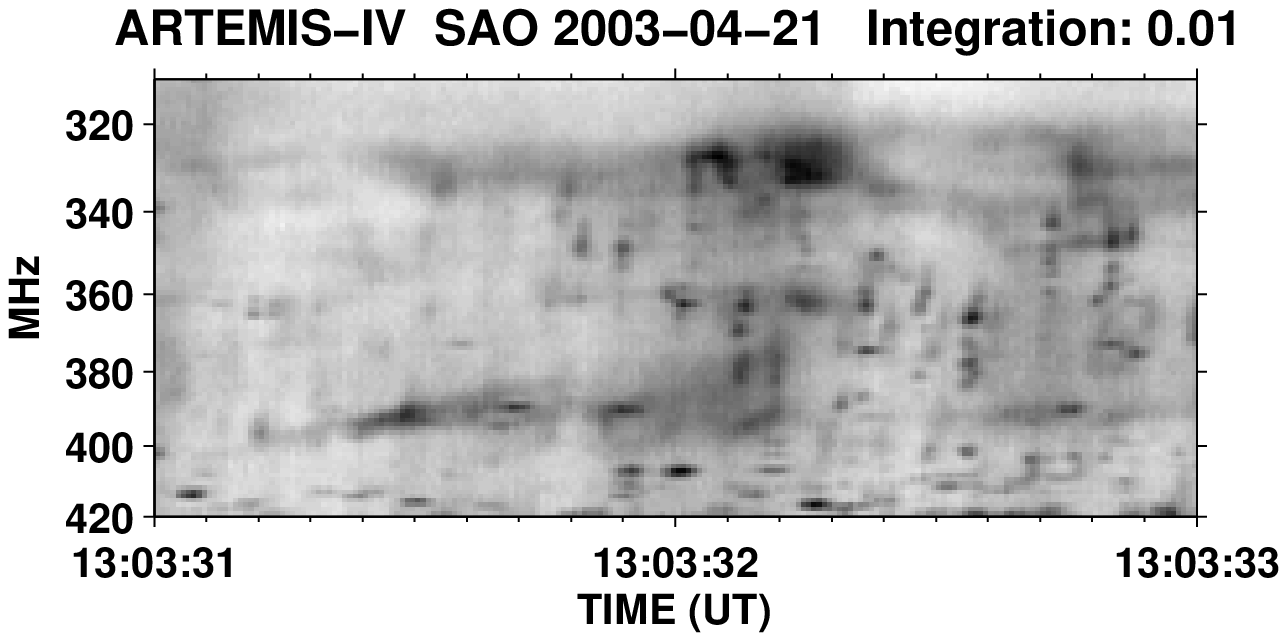} \\ 
\includegraphics[width=\hsize]{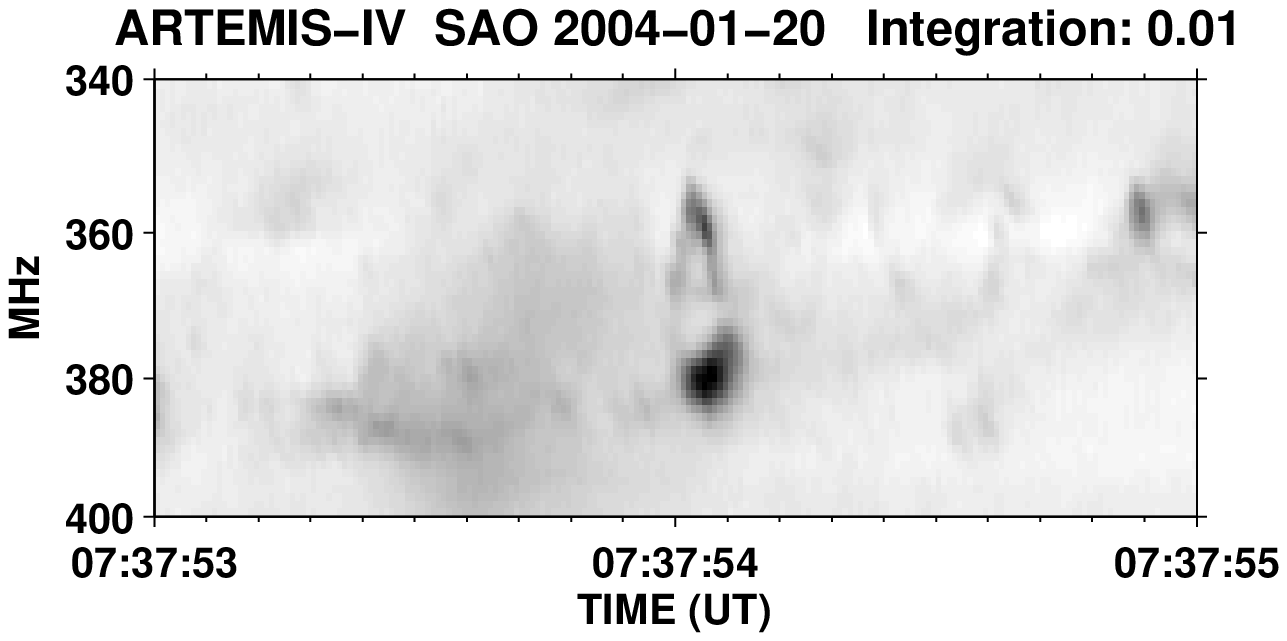}
\caption{Examples of U-like and J-like bursts observed in ARTEMIS/SAO High Resolution (10\,ms) Dynamic Spectra.}
\label{UDot}
\end{center}
\end{figure}

\section{Observations and data analysis} \label{Obs}

\subsection{Instrumentation}\label{Inst}
In this study we used type--IV dynamic spectra recorded by the high sensitivity multichannel acousto-optical analyser (SAO) of the Artemis--IV
solar radio-spectrograph at Thermopylae \citep{Caroubalos01,Kontogeorgos06}. The spectra cover the 270-450 MHz range in 128 channels with a time resolution of 100~samples/sec. The dynamic spectra of the medium-sensitivity, broadband, sweep-frequency receiver ARTEMIS-IV/ASG (650--20 MHz at 10~samples/sec) were used to extend the frequency range.

In assembling the dataset we included all the type-IV meter-wavelength events recorded by both ARTEMIS-IV ASG and SAO during June 30, 1999-August 1, 2010. Thirty-five~metric type--IV events (listed in Table 2 of \cite{Bouratzis2015}, event 21 excluded),  accompanied by well-observed narrow-band structures, were thus selected for further analysis. The associated flares were evenly distributed in heliographic longitude.

A typical event is shown in Fig. \ref{Overall01}. The low time resolution ASG dynamic spectrum depicts a complex event of types III and IV. Embedded within the type-IV continuum (small box in the upper panel of Fig. 1), there is a spike cluster recorded by the SAO receiver with a time resolution of 10ms (lower panel). The GOES light curve indicates that this spike group appeared during the rise phase of the soft X-ray flare.

The Nan\c cay Radioheliograph (NRH) \cite{Kerdraon97} is a synthesis instrument that provides two-dimensional images of the Sun with subsecond time resolution. For the event that we studied in detail (April 21, 2003), the NRH provided data at six frequencies (150.9, 164.0, 236.6, 327.0, 410.5, and 432.0\,MHz) with a cadence of 150 ms. All six frequencies are within the spectral range of the ASG, while the last three are also within the range of the SAO. Radio emission at these frequencies originates in the low and middle corona (0.1-0.5 \RSUN). However, in the case of narrow-band structures originating within flaring regions, the plasma density is significantly enhanced so the above height range is instead a lower limit of the source height.

From the original NRH visibilities, we computed two-dimensional (2D) images with a resolution of 1.13\arcmin\ by 1.57\arcmin\ at 432\,MHz. We also computed one-dimensional (1D) images, using the baselines of the EW and NS antennas with resolutions of 0.6\arcmin\ (EW) and 0.75\arcmin\ (NS),
respectively; the improved resolution occurs because the extension antennas make a very small contribution to the 2D images. We performed self-calibration based on redundant baselines, and this improved the quality of the 1D images significantly.

\section{Results}\label{Res}

\subsection{Individual bursts} \label{SingleBurst}

\subsubsection{Morphology of Individual bursts} \label{Morphology}
The high time resolution of our spectra made the morphological  distinction possible between a number of different narrow-band structures which  on a low resolution recoding would have appeared structureless.  These variants of the standard spike usually appear in the form of type-III family bursts, type J and U, as well as entirely different types,  such as inverted U, reported at decametric wavelengths by \citet{sawant1976microscopic}. The characteristics of these bursts were compared with the standard spikes of our sample and were found well within the average characteristics.

Examples are presented in Fig. \ref{UDot}; they include U--like and J--like  spikes and a reverse U--like spike with an almost simultaneous U--like one. The U--like and J--like spikes have a relative bandwidth of about  3\%  and a total duration 100\,ms. The reverse U--like spike has a relative bandwidth of 2.6\% and total duration of 90\,ms, quite close to the characteristics of the accompanying U-like spike, which were found to be 3.6\% and 50\,ms, respectively.

\subsubsection{Duration and bandwidth of Individual bursts} \label{DurBW}

On the  10 ms SAO dynamic spectra, we measured the duration and the instantaneous spectral width of 11579 narrow-band bursts. The identification of individual bursts was done by inspection, and the width was measured after fitting the temporal and spectral profiles with a smooth curve. 

\begin{figure}
\begin{center}
\includegraphics[height=0.20\textheight,width=\hsize]{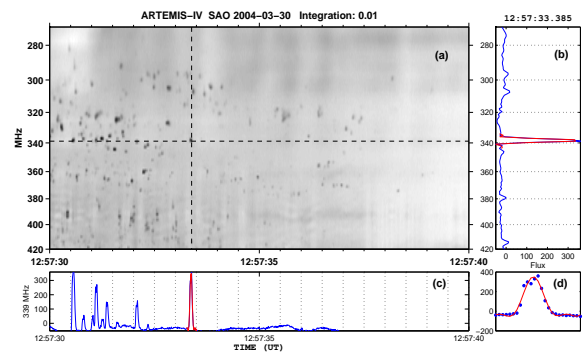}
\caption{Measurement of the width of temporal and frequency profiles of an individual spike within a cluster shown in the dynamic spectrum (a). Panel (b) shows the spectral profile and panel (c) the temporal profile. Panel (d) shows the full temporal profile at 340\,MHz.}
\label{SpikeProfileSpectrum01}
\end{center}
\end{figure}

An example of the measurement of the spectral and temporal width of a spike is presented in Fig. \ref{SpikeProfileSpectrum01}. It shows part of the dynamic spectrum of a spike group and the measurement of the width of temporal and frequency profiles of an individual spike. The side panels show the spectral and temporal profiles of the spike, while  in panel (c) we show the full temporal profile at 340\,MHz.

\begin{figure}
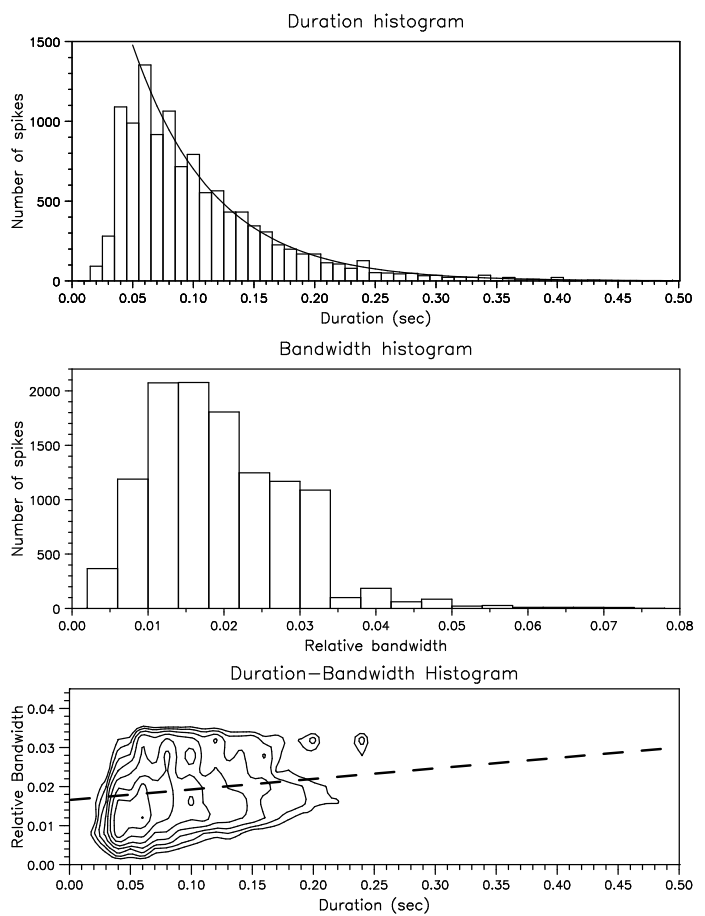

\centering
\includegraphics[angle=-90,width=\hsize]{FIG4A_2.eps} \\
\vspace{.1cm}
\includegraphics[angle=-90,width=\hsize]{FIG4B.eps} \\
\vspace{.1cm}
\includegraphics[angle=-90,width=\hsize]{FIG4C_2.eps} \\
\caption{Top panel: Histogram of spike duration; the line is an exponential fit. Middle panel: Distribution of the relative instantaneous bandwidth of spikes. Bottom Panel: 2D histogram of the number of spikes as a function of relative bandwidth and duration; contours are logarithmically spaced at 99, 70, 49, 34, 24, 17, 12 and 8\% of the peak. The dashed line is the result of  linear regression on the bandwidth-duration scatter plot.} 
\label{SpikeDuration}
\end{figure}

\begin{figure}
 \centering
 \includegraphics[width=\hsize, height=0.15\textheight]{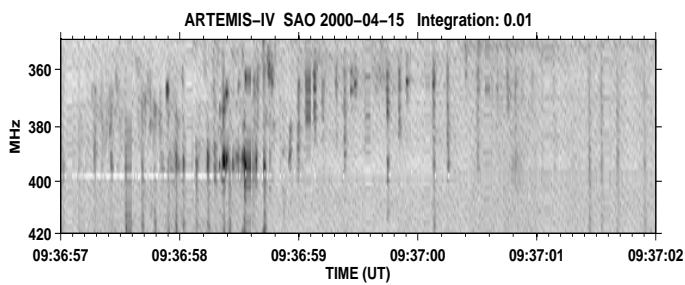}
 \caption{ARTEMIS/SAO Dynamic Spectra of Broadband Spike Bursts.}
 \label{BSBSpikes}
\end{figure}

\begin{figure}[h]
 \centering
\includegraphics[width=0.48\textwidth]{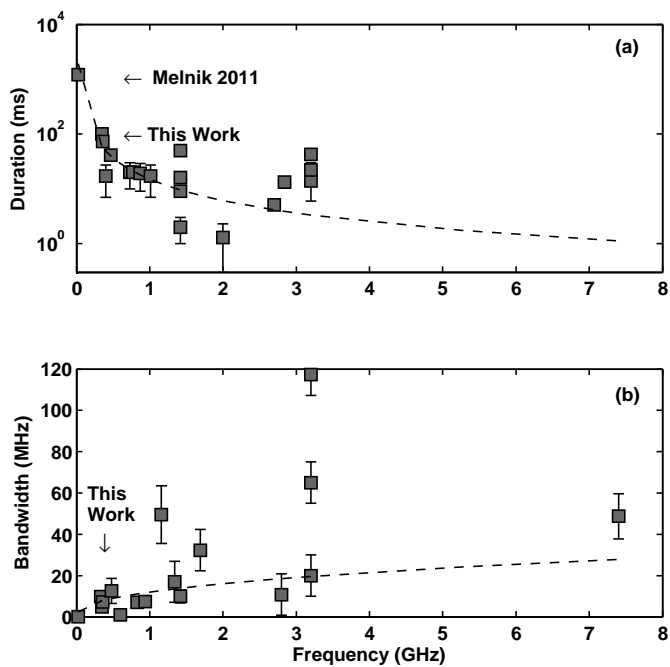}
\caption{ Panel (a): Duration of narrow-band bursts and spikes as a function of frequency of observation. The dashed lines represents the empirical power law fit (${dt \propto {f^{ -1.3}}}$), \citep{Guedel1990, Meszarosova03, Rozhansky2008}. Panel (b):  Instantaneous bandwidth versus frequency; the  line is a plot of the empirical power law fit ${Df \propto 0.66{f^{ 0.42}}}$  by \citet{Csillaghy1993}.}
\label{FreqDurTableGraph}
\end{figure}  

The top panels of {Fig. \ref{SpikeDuration} show histograms of the duration and the bandwidth of the narrow-band structures and the
left column of Table \ref{TableSpikes} summarizes our results. The last column of the table gives the results for for the 2003 April 21 event analyzed in Section \ref{NRH}. The duration histogram is asymmetric with a peak at 60\,ms and a long tail toward high values.{\bf \  The tail is represented well by an exponential decay with a characteristic time of 0.067\,s, while a power law function did not fit the data well}. The true maximum might be an even lower value, owing to the limited instrumental resolution.  The mean value of the distribution was 100\,ms, the median 85\,ms, and the FWHM 135\,ms; 63\% of the spikes had a duration of~$<100$\,ms and  29\% in the 100 to 200\,ms range, and the remaining 8\% had a duration of $>200$\,ms. 

\begin{table*}
\begin{center}
\caption{Average parameters of individual spikes, drifting chains, and spikes, with the last column referring to the 2003 April 21 event.}
\label{TableSpikes}
\begin{tabular}{lcccccc}
\hline
Parameter                               &\multicolumn{1}{c}{Individual} &\multicolumn{2}{c}{Direct}     &\multicolumn{2}{c}{Reverse} &\multicolumn{1}{c}{Event of}\\
                                                &               & Chains                & Columns & Chains                & Columns       &2003-21-04     \\
\hline
\hline
Number of events                        &11579                          & 63              & 5                     &       5               &       9               &               \\
Duration (s)                            & 0.1~                          &   5.8           & 0.04          &       1.29    &       0.05    & 0.048         \\
Bandwidth (MHz)                 & 7.8                           & 24.6          & 21.0            &       7.4             &       24.9    & 7.7           \\
Relative bandwidth              & 0.02                          &  0.070                &  0.054          &       0.019   &       0.066   & 0.021         \\
Drift rate (MHz/s, 295 spikes)  &$-390$,~~ 506                          & $-$6.8  & $-$1260       &       8.1             &       517             & $-1800$ \\
Log drift rate (s$^{-1}$)       &$-1.12$,~~ 1.43                        & $-$0.019        & $-$3.25       &       0.021   &       1.36    & $-4$          \\
\hline
\end{tabular}
\end{center}
\end{table*}

The average relative bandwidth of the spikes, $\delta \ln f$, was 2\%, the median 1.8\% and the FWHM 2.6\%. We note that 99\% of the spikes had $\delta \ln f < 5$\%. Only 0.37\% of the total spike events had a relative bandwidth of $ > 7$\% and these can be considered as outliers. The mean duration of these outliers was 150\,ms, twice as large as that of narrow-band spikes. In one case we observed a group of very large bandwidth spikes with $\delta \ln f \approx 20\%$  (Fig. \ref{BSBSpikes}); this is similar to that of the large spikes reported by \citet{Bakunin85} in a different  frequency range  (175--235\,MHz).

The bottom panel of {Fig. \ref{SpikeDuration} shows a 2D histogram of the number of spikes as a function of relative bandwidth and duration. The peak is at $\delta \tau \approx 60$\,ms and $\delta \ln f \approx 12$\%. Secondary peaks are probably not significant, since their separation is not greater than two histogram bins ({\it cf.} duration histogram); thus we cannot consider them as evidence for separate populations in the distribution. The contour plot shows a sharp drop of bandwidth above 3.5\% and a tendency by the bandwidth to increase with duration. A linear regression on the bandwidth-duration scatter plot, limited to bandwidths of less than 3\%, gave  a slope of {\bf $d (\delta \ln f) / d(\delta \tau) =0.027\pm0.0.001$\,s$^{-1}$ (dashed line in the figure); we note, however, that the lowest three contours at small bandwidths gave a steeper slope} of $0.083\pm0.004$\,s$^{-1}$.




The duration--frequency dependence is usually expressed by a phenomenological power law of the form $ D \sim f^{-\alpha}$. Our results are consistent with the empirical relation proposed by \cite{Guedel1990}, \cite{Meszarosova03}, and \cite{Rozhansky2008}.
In Fig. \ref{FreqDurTableGraph}(a) we plot  data from Table 1 of \citet{Dabrowski2011}, together with our results and those of \citet{Melnik2011, Melnik2014}; the curve is a power law with $\alpha=1.32,$ which is the average  of the power reported by \citet{Guedel1990} ($\alpha=1.34$) and by \cite{Meszarosova03} and \cite{Rozhansky2008} ($\alpha=1.29$). This plot indicates a reasonable  power law fit up to  $\sim~2$\,GHz. There appears to be a decrease in duration with frequency from 270 MHz to 2 GHz and then an increase above that range. If  this is real, it might indicate a change of emission mechanism in this spectral region.

The instantaneous relative bandwidth--frequency relationship is expressed by the empirical power law $\delta \ln f \sim 0.66f^{ 0.42}$  \citep{Csillaghy1993}.  In Fig. \ref{FreqDurTableGraph}b we overplot this power law on the bandwidth-frequency data set used for Fig. \ref{FreqDurTableGraph}a. Our measurement is near the value expected from the empirical relation.

\begin{figure*}
 \centering
\begin{tabular}{cc}
\multicolumn{2}{c}{\includegraphics[trim=0cm 0.5cm  0cm 0cm,clip,width=0.72\textwidth]{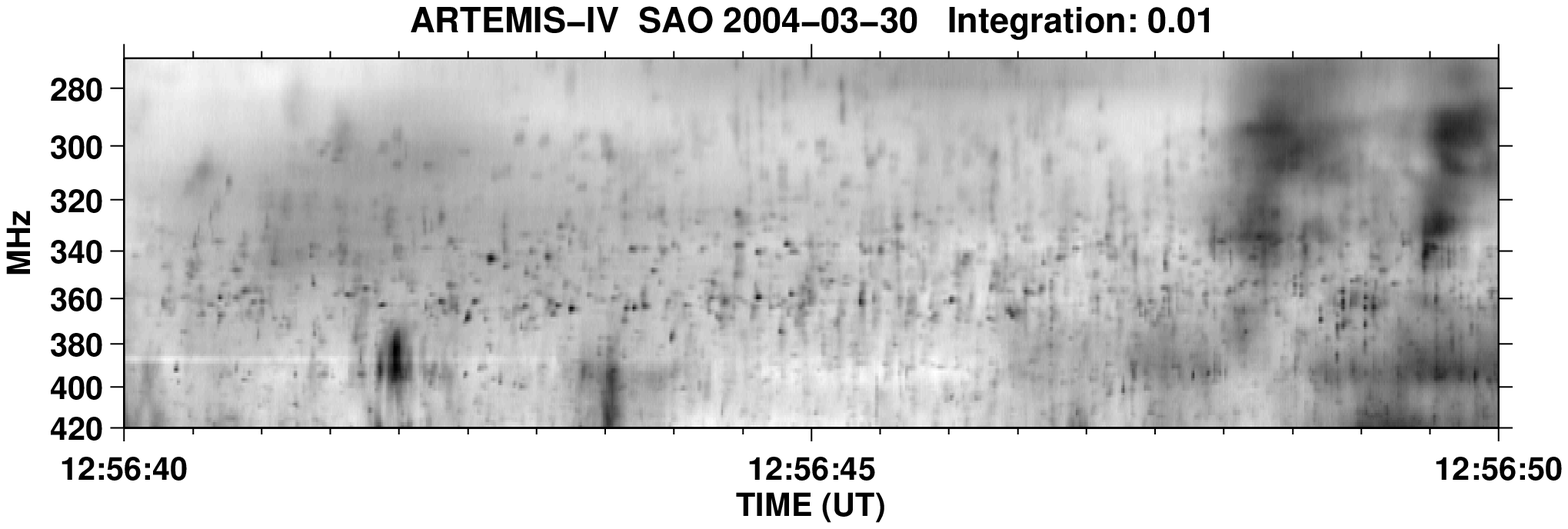}} \\
\multicolumn{2}{c}{(a)} \\
\includegraphics[trim=0cm 0.5cm  0.2cm 0cm,clip,width=0.36\textwidth]{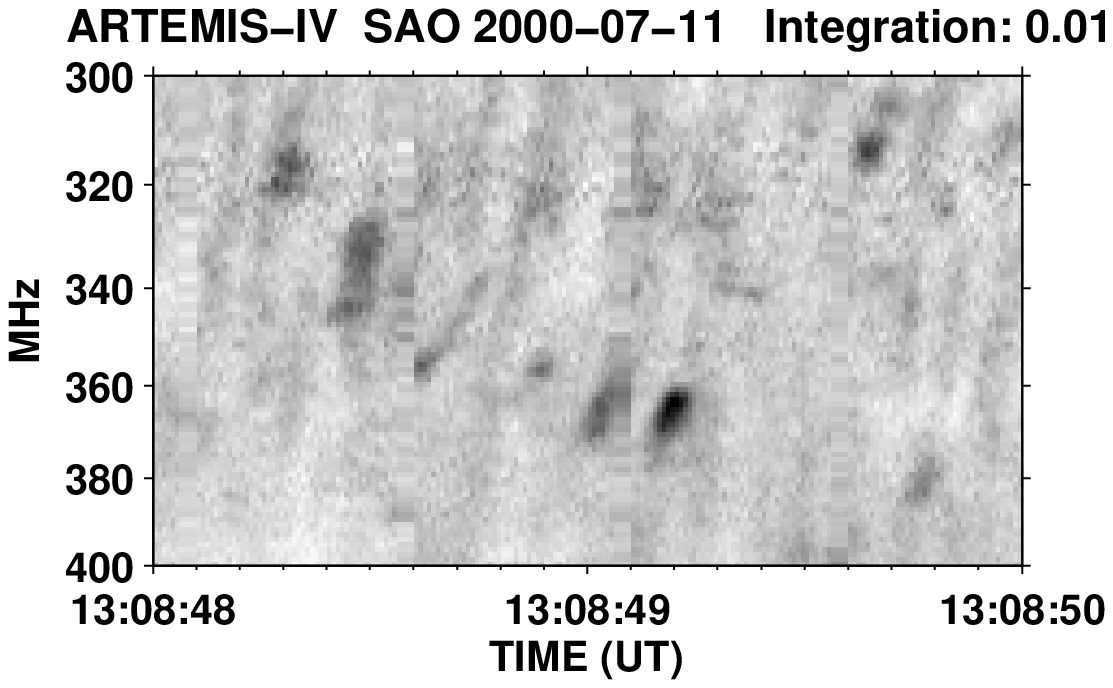} & 
\includegraphics[trim=0cm 0.5cm  0.2cm 0cm,clip,width=0.36\textwidth]{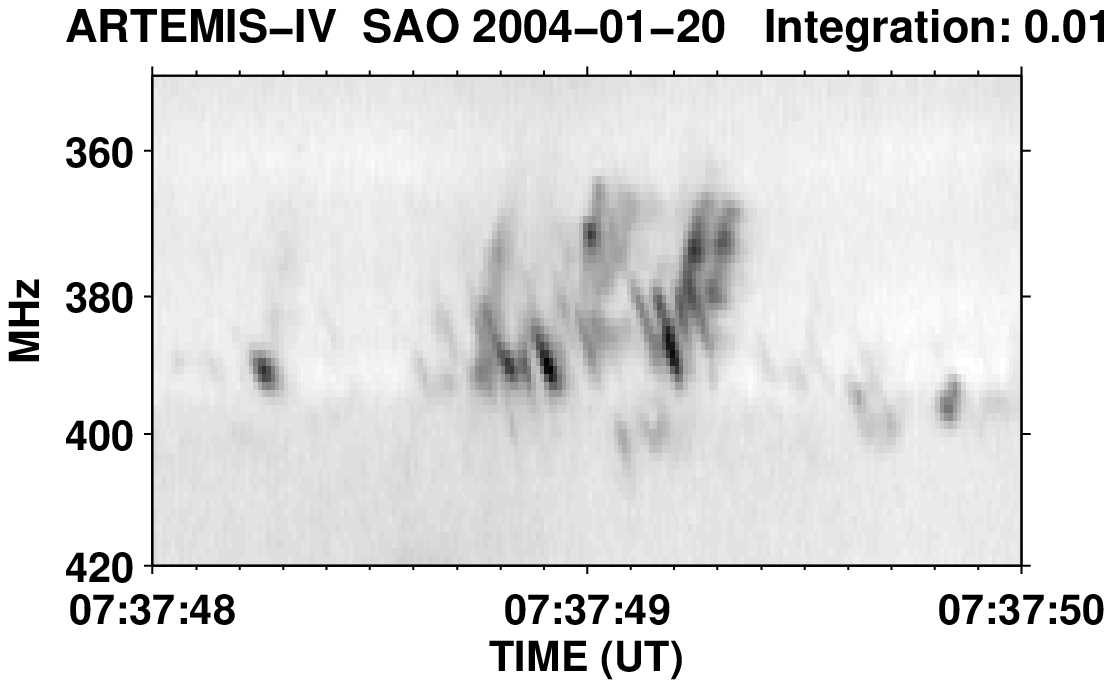}\\
(b) & (c) \\
\multicolumn{2}{c}{\includegraphics[width=0.7\textwidth]{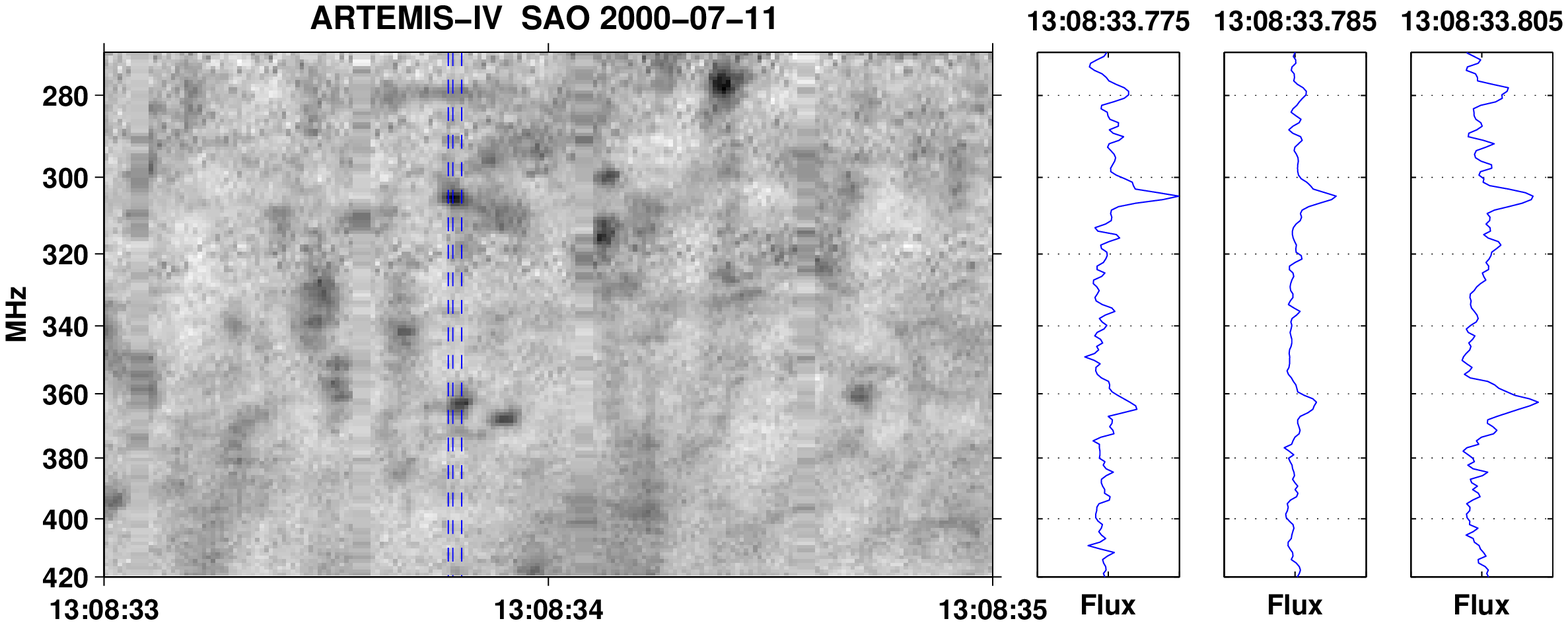}} \\
\multicolumn{2}{c}{(d)} \\
\end{tabular}
\caption{ARTEMIS/SAO high resolution (10\,ms) dynamic spectra of drifting narrow-band structures. (a) Spikes with different values of drift rates: positive, negative, and unmeasurable. (b) Spikes with negative drifts, (c) spikes with positive drifts, and (d) spikes with unmeasurable drift rate. Plots in (d) are spectral cuts along the lines marked in the dynamic spectrum}
\label{DriftingSpikes}
\end{figure*}

\subsubsection{Frequency drift of individual bursts} \label{FrequencyDriftRate}

Some spikes show frequency drifts that are both positive and negative (Fig. \ref{DriftingSpikes}). Assuming a negligible intrinsic bandwidth of the emission, there is an upper limit on the value of the drift that can be measured, depending on the observed bandwidth of the structure, $\delta \ln f$, and on the accuracy of measurement of the value of a maximum in time, $\delta t$, which is roughly equal to $\delta \ln f /\delta t$.  For a relative bandwidth of 2\% (see previous section) and $\delta t\sim0.005$\,s (half the SAO time resolution), this gives a value of $\sim4$\,s$^{-1}$ as an approximate upper limit to the measurable relative drift rate.

We made indicative drift measurements for 295 spikes. The measured values of $d\ln f/dt$ ranged from $-3.5$ to 4.5\,sec$^{-1}$, roughly within the limits computed in the previous paragraph. Most spikes (60\%) showed negative drifts with an average value of $-390$\,MHz\,s$^{-1}$, corresponding to a logarithmic drift of $-1.12$\,s$^{-1}$ (Table \ref{TableSpikes}).
This should be compared to the type-III frequency drift rate, which ranges from  about $-0.42$ to $-1.0$\,s$^{-1}$ \citep{1978BenzZlobec,benz2009b}. We note, moreover, that spike-associated type IIIs have a lower drift rate \citep[see Table 1 of][ from which $d\ln f/dt\simeq-0.35$\,s$^{-1}$]{Benz1996}.

Fewer spikes (21\%) showed a positive drift with an average of $506$\,MHz\,s$^{-1}$ and a corresponding logarithmic drift  of $1.43$\,s$^{-1}$, while 19\% of the cases showed no measurable drift. The average drift over all spikes was $-145$\,MHz\,s$^{-1}$, and the logarithmic drift was $-0.47$\,s$^{-1}$.

\begin{figure}
\includegraphics[width=0.40\textwidth]{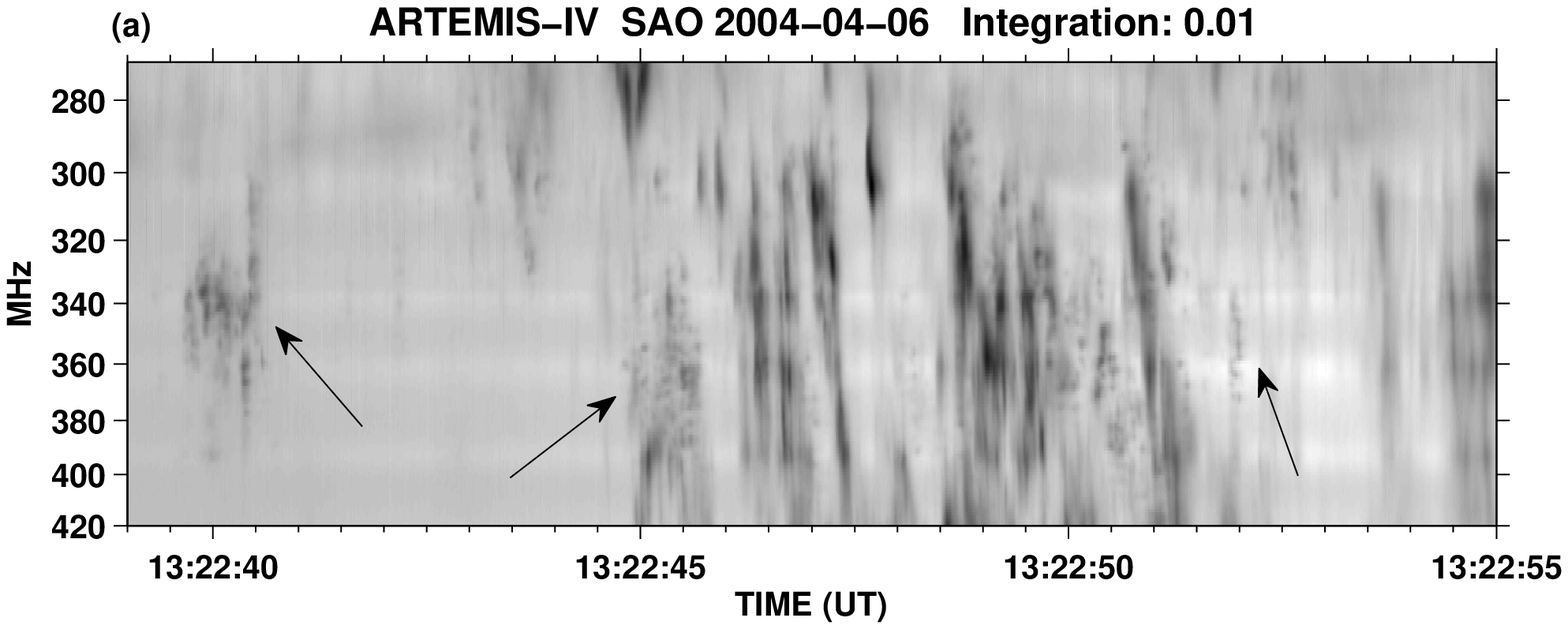}\\
\includegraphics[width=0.40\textwidth]{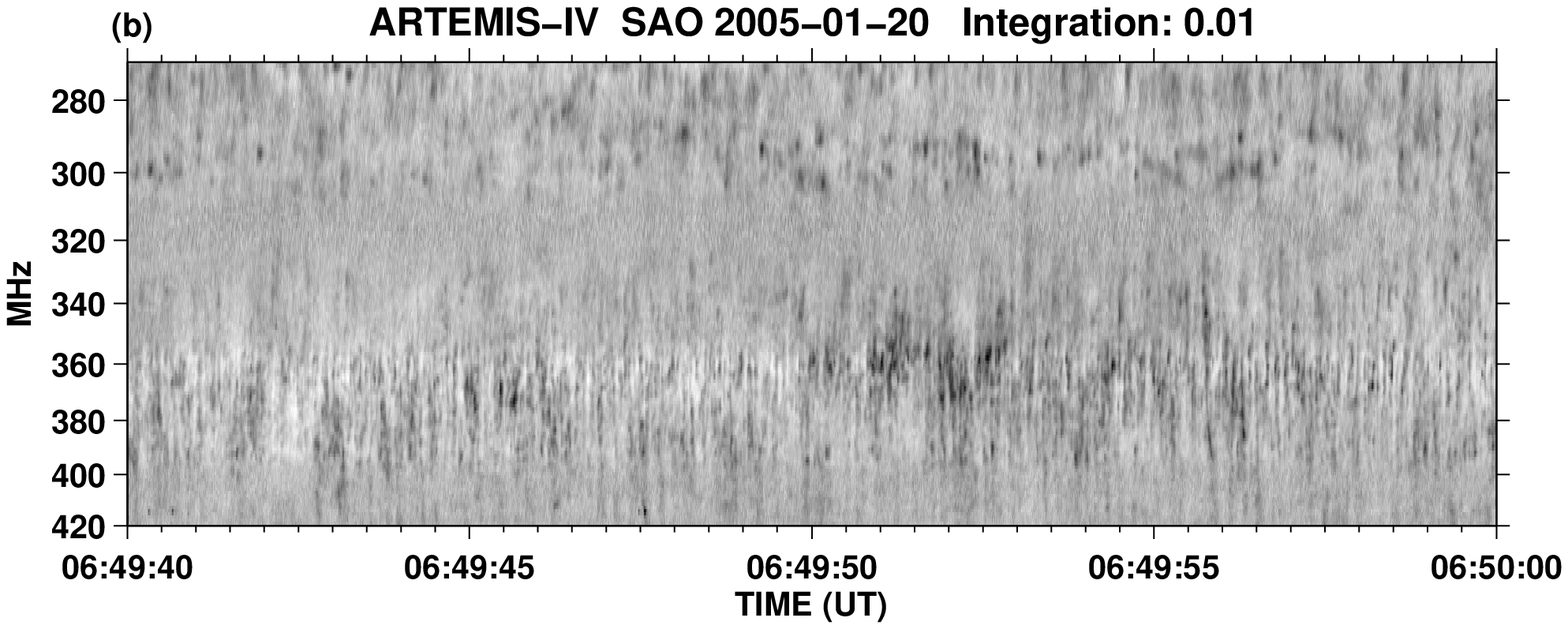}
\caption{ARTEMIS/SAO high resolution (10\,ms) dynamic spectra. (a) Spike groups in time. The arrows indicate some of these clusters. (b) Spike groups in frequency. There are two lanes of spikes in dynamic spectrum with 300 and 380 MHz central frequencies.}
\label{GroupInTime}
\end{figure}

From the frequency drift rate, $d\ln f/dt$, and the relative bandwidth, $\delta \ln f$, we may estimate the exciter speed, $V_{exc}$, and the vertical source size, $\Delta R$, by multiplying by the ambient coronal plasma frequency scale height, $H_f=(d\ln f\,dR)^{-1}$: 
\begin{eqnarray}
V_{exc}=\frac{dR}{dt}&=&\frac{d\ln f}{dt}\, H_f\\
\Delta R &=&\delta \ln f~H_f
,\end{eqnarray}
where the frequency scale height is twice the density scale height. This calculation is affected by ambiguities introduced by variations in the ambient medium properties and the model selection \citep[see, e.g.,~][~for a detailed discussion of model selection]{Pohjolainen07, Pohjolainen08}. It may, however, adequately provide the appropriate range of speeds and sizes.

Two well-established quiet-Sun models were used in our calculations. For the twofold \citet{Newkirk} model, $H_f=140\times10^3$\,km for emission at the fundamental and $190\times10^3$\,km at the harmonic, while for the  hybrid model of \citet {Vrsnak04}, $H_f=100\times10^3$\,km for the fundamental and $150\times10^3$\,km for the harmonic. The frequency scale heights are almost constant within the SAO frequency range. Taking the average of the two models, the average drifts quoted above give exciter speeds of $-0.45$c and 0.58c for negative and positive drift spikes, if the emission is at the fundamental. For harmonic emission, the corresponding values are $-0.64$c and 0.82c. 

We note that, in order to keep the exciter speed less than the speed of light, the drift should not exceed 2.5\,s$^{-1}$ for emission at the fundamental and 1.76\,s$^{-1}$ for harmonic emission and that some spikes showed drifts beyond these limits.
If such drifts are interpreted in terms of exciter motion, the local scale height has to be less than predicted by quiet-Sun models. Indeed, this is expected to be the case within disturbed structures such as interacting plasmoids, as proposed by \citet{Dabrowsky2015}, where density gradients are higher.

We also note that, as pointed out previously \citep[e.g.,][]{Poquerusse94,Klassen03b,benz2009b}, the geometric effects make the observed frequency drift rate higher than the true drift by a factor of  $[1-(V_{exc}/c)\cos\theta]^{-1}$, where $\theta$ is the angle between the exciter path and the line of sight. Thus for an exciter moving in the radial direction this factor is unity at the solar limb, but at the center of the disk and for an exciter speed of $0.7c$, it is $\sim 3$; this may have affected some of our measurements.

\begin{figure} 
\centering
\includegraphics[height=0.20\textheight,width=0.45\textwidth]{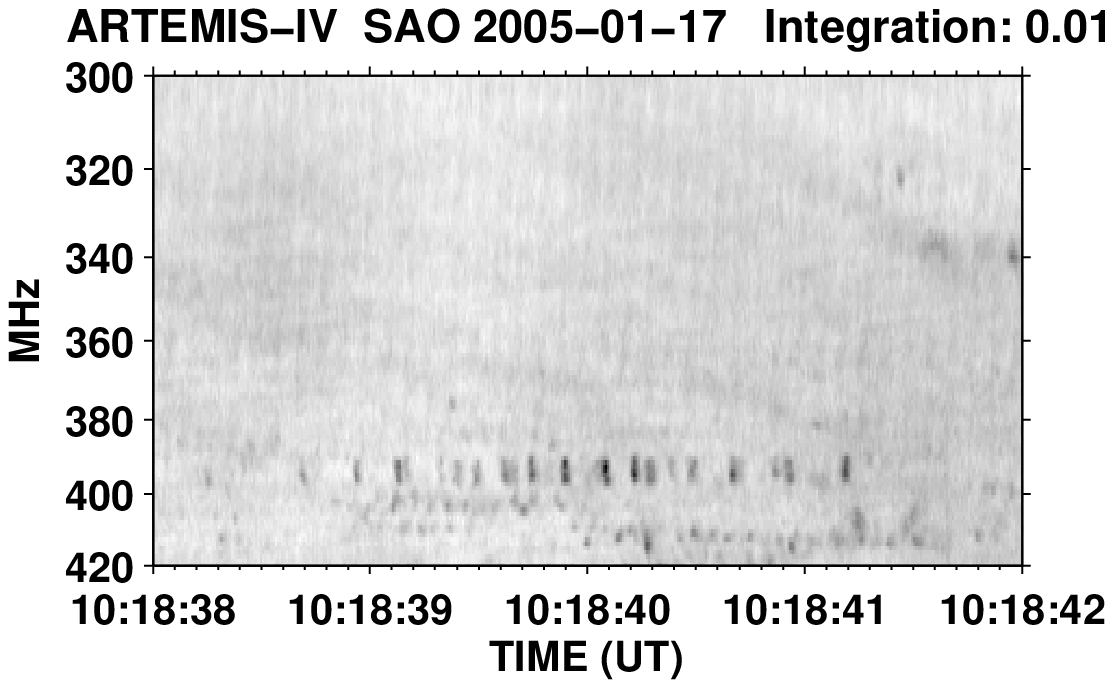}\\
\includegraphics[height=0.20\textheight,width=0.45\textwidth]{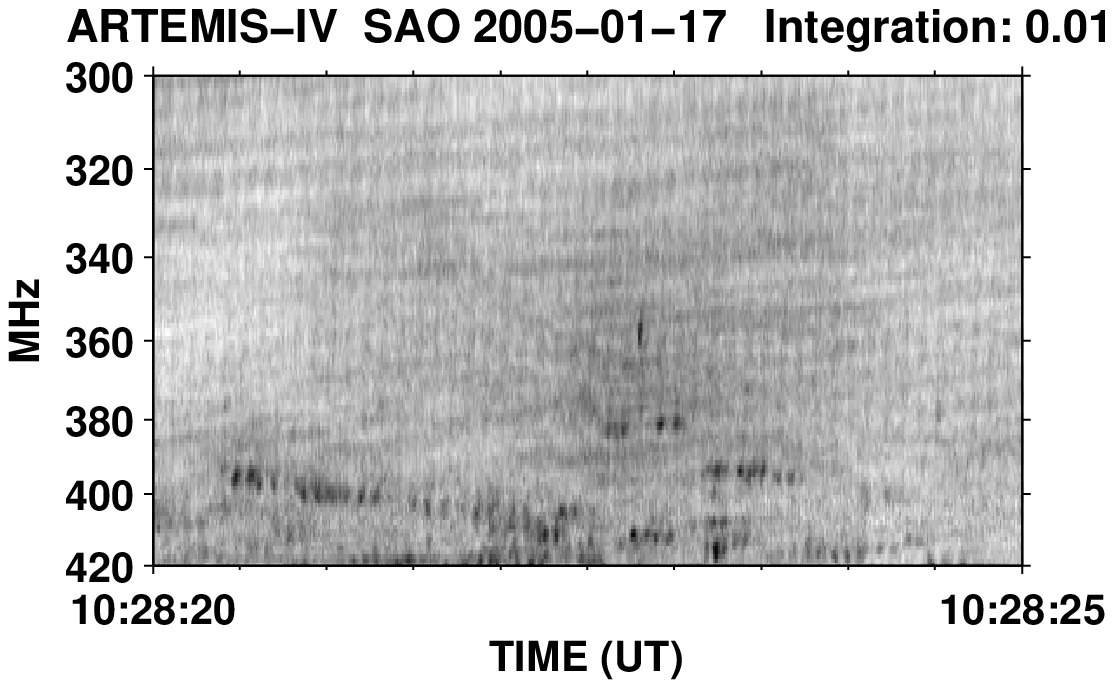} \\
\includegraphics[height=0.20\textheight,width=0.45\textwidth]{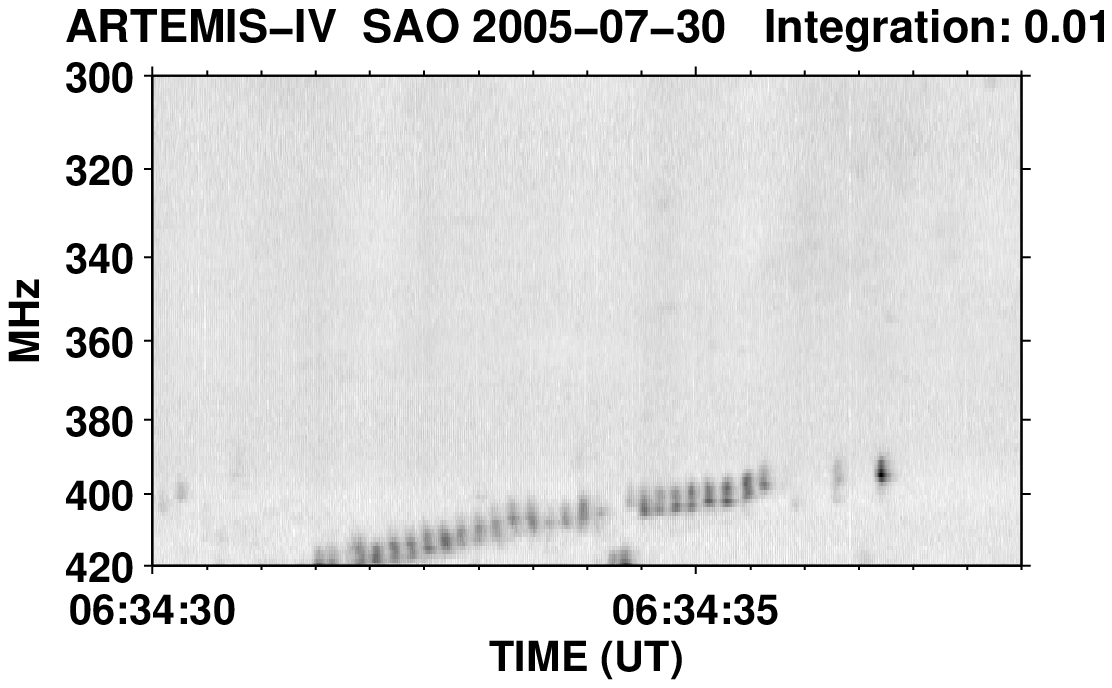} \\ 
\includegraphics[height=0.20\textheight,width=0.45\textwidth]{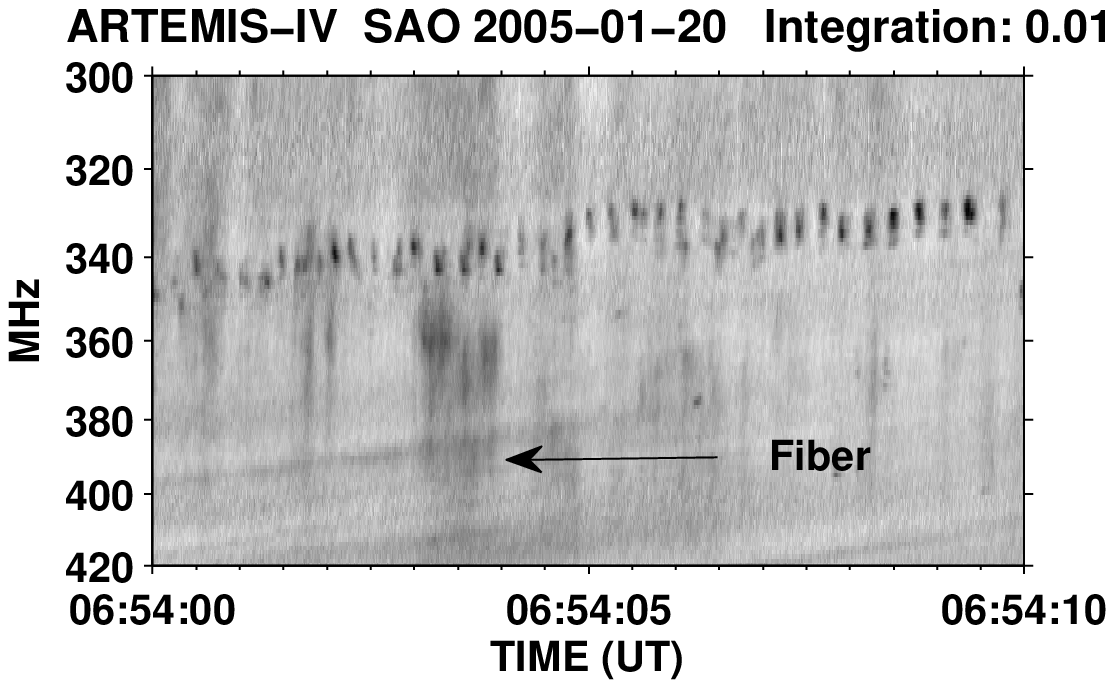}
\caption{ARTEMIS/SAO high resolution (10\,ms) dynamic spectra of spike chains. From top to bottom: Spike chain with no drift. Positive drifting spike chain. Negative drifting spike chains. In the bottom panel there is a fiber that drifts parallel to spike chain.}
\label{SpikeChains}
\end{figure}
 
As for the vertical size of spikes, Equation (2) gives $\Delta R \sim 2.4\times10^3$\,km for emission at the fundamental and  $\sim 3.4\times10^3$\,km for emission at the harmonic for the average bandwidth of 2\%. These are upper limits if the actual density gradient is greater than that of the quiet Sun.

\subsection{Groups and chains of  bursts} \label{ChainBurst}

More often than not, spike bursts are not randomly scattered in dynamic spectra, but are grouped in clusters close in time (Fig. \ref{GroupInTime}a) or in frequency (Fig. \ref{GroupInTime}b).
A particular class of clusters are the {\it columns} \citep{Dabrowski2007}, which are large groups of individual spikes clustered within a very short time interval ($\lesssim1$\,s) over a broad frequency range. We found that negative drifting column spikes had a typical group drift rate of $\sim-1260$\,MHz\,s$^{-1}$, corresponding to a logarithmic drift rate of $d\ln f/dt\approx-3.25$\,s$^{-1}$. This high drift is  similar to that of type-IIId bursts, which are members of the type-III family with very fast drift of about $-1500$\,MHz\,s$^{-1}$ in the 500-100 MHz frequency range \citep{Poquerusse94}. We also note that \citet{Sawant2002} report drift rates between 180 and 1200 MHz\,s$^{-1}$ in the frequency range 1000 to 2000\,MHz for negative drifting, dot-like structures. Positive drifting columns had $d\ln f/dt\approx1.36$\,s$^{-1}$ ($517$\,MHz\,s$^{-1}$).

Another class of interest are the {\it chains} (Fig. \ref{SpikeChains}), most of which exhibited group frequency drift. tTe majority of them exhibited negative drift $d\ln f/dt\approx~-0.021$, while a few drifted toward higher frequencies at a rate  of $d\ln f/dt\approx~0.033 s^{-1}$. The average chain duration in our data set was within the 2--20 s range. The chain drift rate corresponds to the drift of intermediate drift bursts  recorded  at the same time as the chain; an example is presented in Fig. \ref{SpikeChains}(d).

\begin{figure}
\centering
\includegraphics[width=0.5\textwidth,height=0.15\textheight]{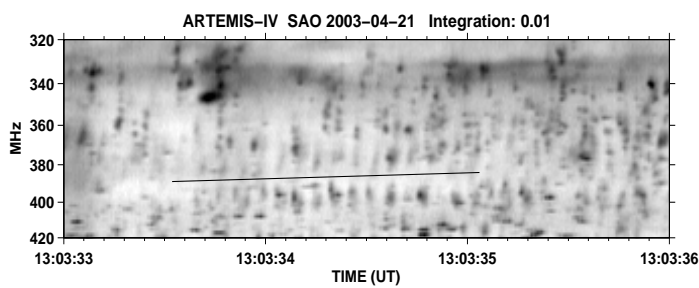}
\caption{ARTEMIS/SAO Dynamic Spectrum with group of bi-directional spikes. Above the line there are negative drifting spikes and below are positive drifting spikes.}
\label{Bidirectional}
\end{figure}

\begin{figure}
\centering
\includegraphics[trim=0cm 0.5cm  0cm 0cm,clip,width=0.45\textwidth,height=0.15\textheight]{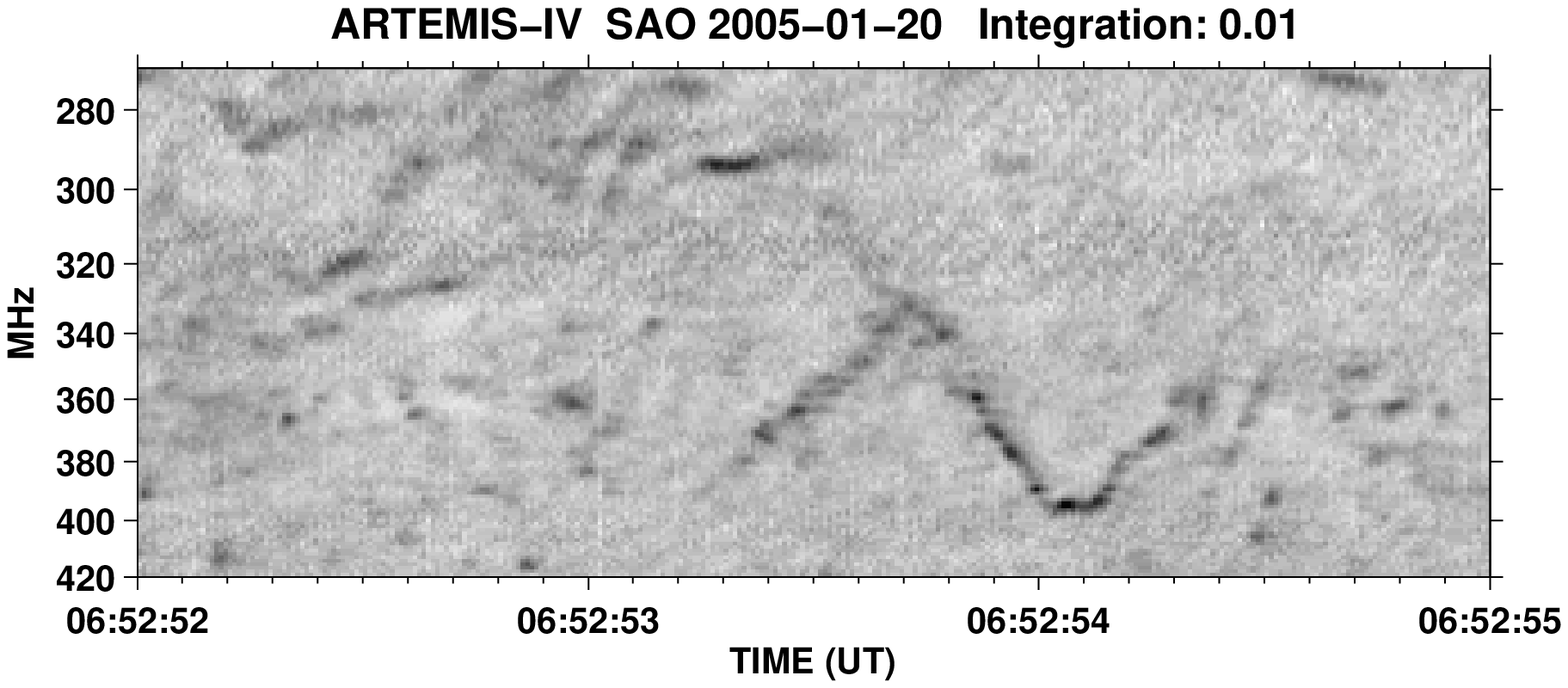}
\includegraphics[trim=0cm 0.0cm  0cm 0cm,clip,width=0.45\textwidth,height=0.15\textheight]{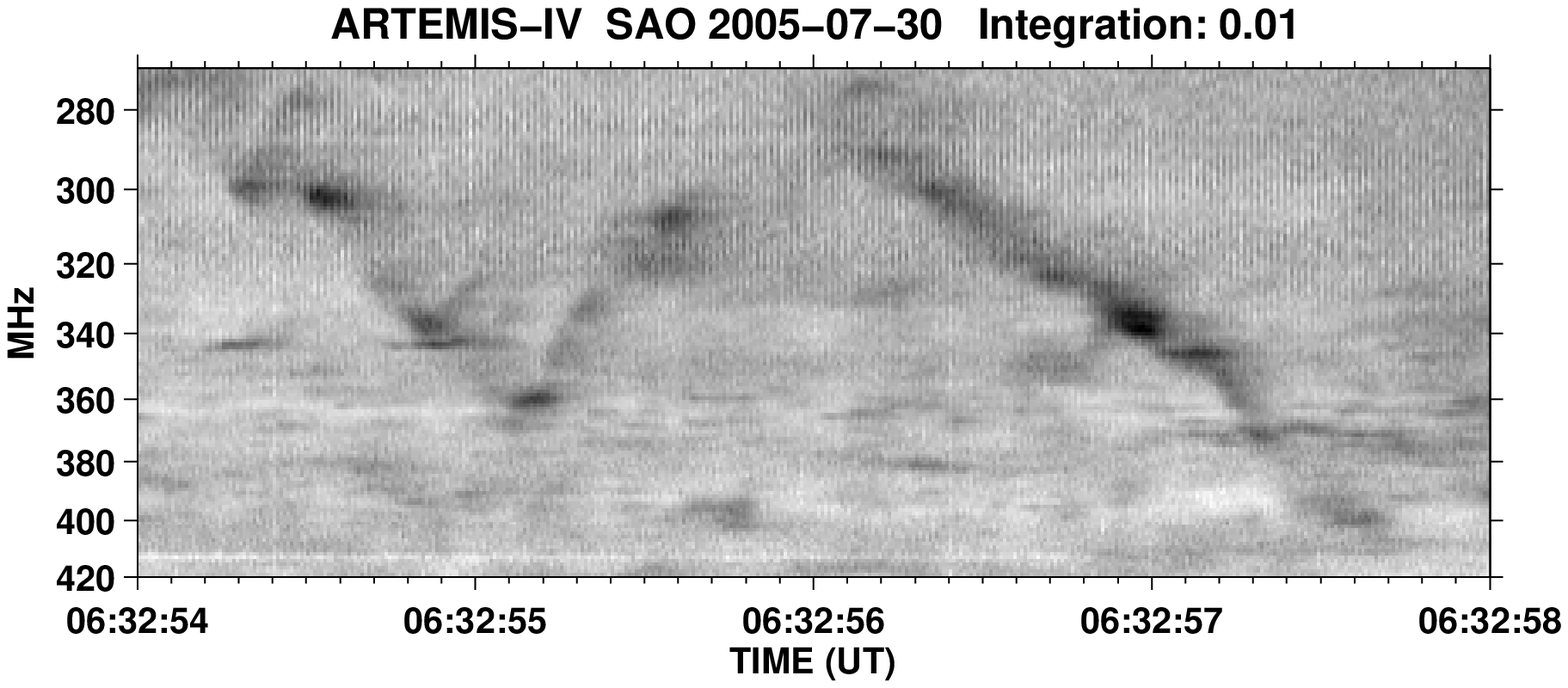}
\caption{High resolution dynamic spectra of spike groups with N-like morphology (see text).}
\label{NSpikes}
\end{figure}

Finally, we found parallel chains, where the spikes of the low frequency chain had negative drift while  the corresponding spikes of the high frequency chain had positive  drift (Fig. \ref{Bidirectional}). These we have dubbed \emph{Bi-directional Spikes,} and they  probably trace small scale X--reconnection events.

\subsubsection{Peculiar spikes and groups}
ARTEMIS-IV recorded some peculiar and very rare kinds of spike groups. These kinds of structures form patterns in dynamic spectra such as sequences of spikes arranged in the form of N bursts \citep{Caroubalos1987} or in the form of lace bursts \citep{Karlicky01}.

\begin{figure}[t]
\centering
\includegraphics[width=0.45\textwidth,height=0.15\textheight]{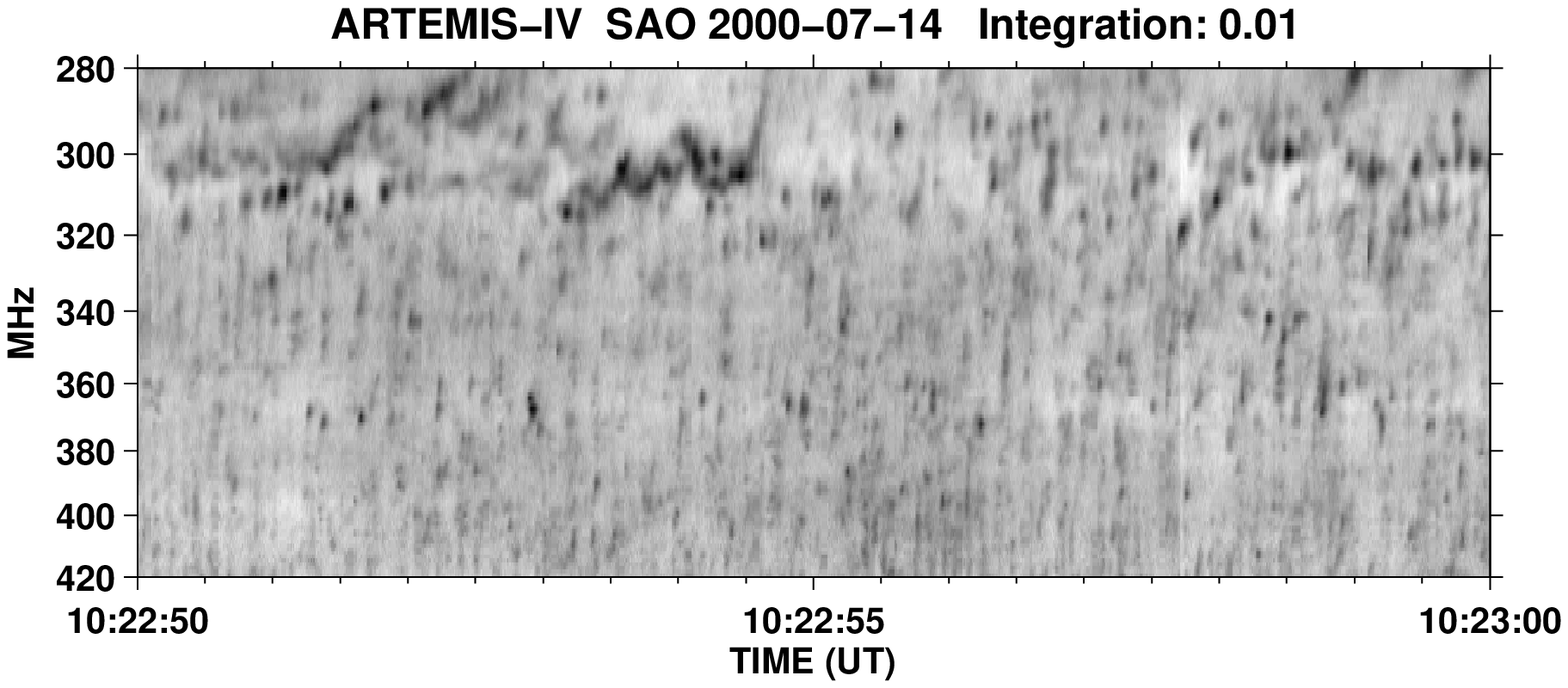}
\includegraphics[width=0.45\textwidth,height=0.15\textheight]{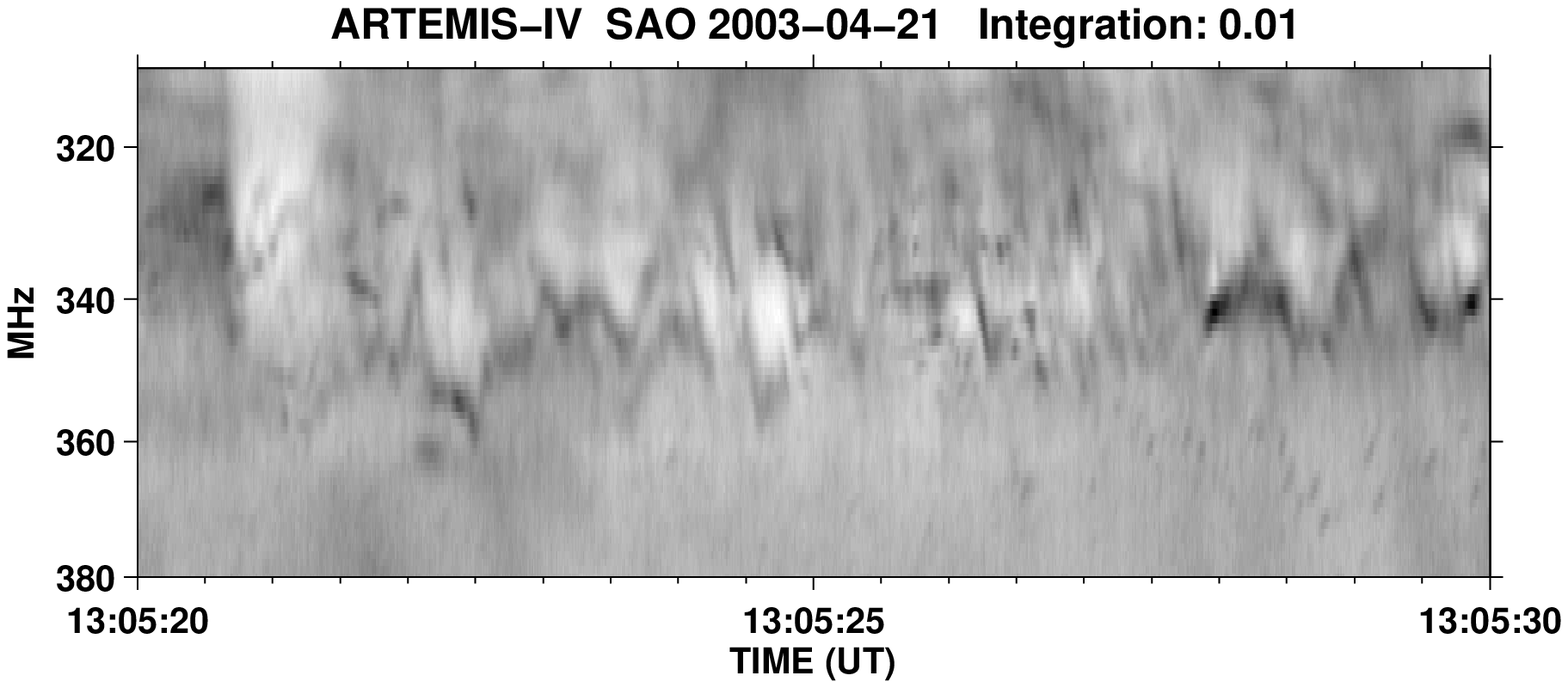} 
\caption{Dynamic spectra of spike groups exhibiting lace-like morphology (see text).}
\label{LaceSpikes}
\end{figure}

Figure \ref{NSpikes} gives examples  of two N burst-like patterns that consist of spikes. The duration of this kind of structure is about  1-3 seconds. The relative drift rate of each component is on the order of 0.4\,s$^{-1}$, which is comparable to type-III bursts' drift rate. We do not classify these peculiar groups as chains because there are three components of spike chains in a row. Furthermore, the drift rate of fiber-like spike chains is less than the drift rate of each component. Sometimes, a group of spikes form a lace-like pattern (Fig. \ref{LaceSpikes}).

Similar results for superfine structures of zebra and fiber bursts consisting of spikes in the 5.2--7.6 GHz frequency range were reported by \citet{Chernov2012}  and \citet{Kuznetsov2007} for two events on May 29, 2003 and April 21, 2002, respectively, and by \citet{2012_TanBaolin&al} for zebra bursts in the microwave frequency range.
The superfine structures of zebra and laces have been theoretically studied by \citet{2001_Barta&Karlicky}, \citet{2005_Barta&Karlicky_b} and \citet{2011_Barta&al_a}. They proposed the double resonance emission mechanism at points where the upper hybrid plasma frequency equals a low harmonic of the electron cyclotron frequency for the lace and zebra emission bands. In a turbulent background, the cascade to small spatial scales accounts for the spike superfine structure.

\section{Spatially resolved structures}\label{NRH}
Most works on spatially resolved spikes \citep{Krucker1995, Krucker1997, Paesold2001, Benz2002} refer to type-III-associated spikes, whereas \cite{Battaglia2006} report spikes apparently embedded in  a type-IV continuum. In all cases the emphasis was on the displacement of the spike sources from the soft X-ray flare by more than100\arcsec. In this section we present results from the analysis of simultaneous SAO/NRH observations of spikes during the 2003
April 21 event.

\begin{figure}[]
\centering
\includegraphics[width=\hsize]{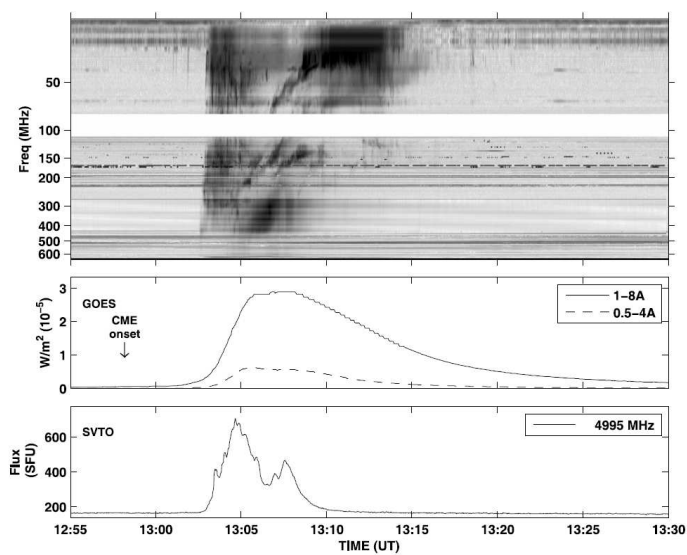}
\caption{Dynamic spectrum of the 2003 April 21 event, recorded with ARTEMIS-IV (combined ASG/SAO data). Time profiles of the soft X-ray emission (GOES) and the microwave emission at 4995\,MHz (San Vito) are given for reference. Figure adapted  from \cite{Bouratzis2015}. }
\label{overview}
\end{figure}

\subsection{Overview}
The selected event was a complex one associated with a GOES M2.8 class flare and a CME. It was rich in type IIIs during the rise phase of the microwave emission (Fig. \ref{overview}), which extended into interplanetary space (see http://secchirh.obspm.fr/survey.php). It also had two type-II bursts with a fundamental-harmonic structure, a type-IV continuum, and a moving type-IV. The corresponding flare occurred in NOAA active region 10338 near the center of the solar disk at N18E02. 

\begin{figure}[]
\centering
\includegraphics[width=\hsize]{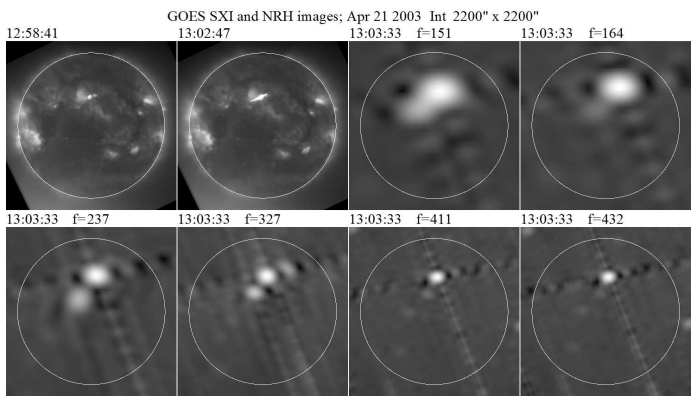}
\caption{GOES/SXI images before and during the early phase of the flare and NRH images averaged over 23\,s during an interval of spike activity. The white circle marks the photospheric limb. {\bf All images are oriented in the solar E-W, N-S direction.}}
\label{sxinrh}
\end{figure}

\begin{table}[]
\begin{center}
\caption{Parameters of the {\bf strongest} background source and the NRH beam, 2003 April 21 event. Positions and sizes are in \arcsec}\label{2003a}
\begin{tabular}{lrrrrrrr}
\hline
Parameter                       &\multicolumn{1}{c}{SXR}        &\multicolumn{6}{c}{NRH frequency (MHz)}\\
                                        &               & 151   & 164   & 237     & 327   & 411   & 432   \\
\hline
$T_b$, 10$^8$\,K        &               & 51    & 128   & 17    &       2.1     &2.6    & 2.7     \\
Position, EW                    & $-21$ & 54    & 176   & 68    & 50    & 34      & 34    \\
Position, NS                    & 370   & 428   & 498   & 474   & 464   & 448     & 448   \\
B$_{maj}$                       &               & 409   & 317   & 209   & 164     & 131   & 125   \\
\smallskip
B$_{min}$                       &               & 271   & 230   & 164   & 142     & 104   & 101   \\
2D beam\\
\hline
B$_{maj}$                       &               & 270   & 249   & 173   & 125     & 99    &  94   \\
\smallskip
B$_{min}$                       &               & 194   & 178   & 123   & 89      & 71    &  68   \\
1D beam\\
\hline
B$_{NS}$                        &               & 194   & 178   & 123   & 59      & 47    &  45   \\
B$_{EW}$                        &               & 270   & 249   & 173   & 48      & 38    &  36   \\
\hline
\end{tabular}
\end{center}
\end{table}

Intense spike emission was observed during the early phase of the event with a duration of about 23\,s from 13:03:22 to 13:03:45~UT.~Fig. \ref{sxinrh} shows NRH images at 150.9, 164.0, 236.6, 327.0, 410.5, and 432.0\,MHz, averaged over this time interval. The same figure shows two GOES/SXI images before and in the early phase of the event at 12:58:41 and 13:02:47~UT. At high frequencies, a single source with simple structure is seen, whereas we have three sources at 327\,MHz and two sources at lower frequencies. The main radio source was displaced by about 100\arcsec\ NW with respect to the soft X-ray flare at 432\,MHz and about 240\arcsec\ at 164\,MHz. The position and the size of the {\bf strongest} background source, together with the instrumental resolution are given in Table \ref{2003a}. In the same table we give the NRH resolution for each frequency. The source size is not  much greater than the beam size, so the actual size is smaller and the brightness temperature higher than observed.

\begin{figure}
\centering
\includegraphics[width=\hsize]{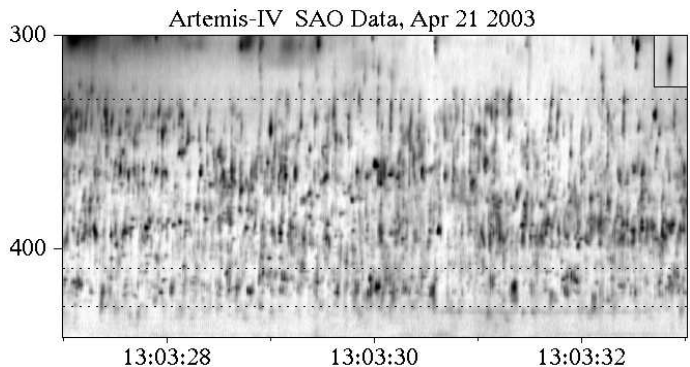}
\caption{SAO dynamic spectrum of spikes during a six-second interval at full time resolution (10 ms). Dotted lines mark the frequencies of NRH data at 327.0, 410.5, and 432.0\,MHz. The insert in the top right corner shows the autocorrelation function.}
\label{spikes2}
\end{figure}

\begin{figure*}
\centering
\includegraphics[width=.70\textwidth]{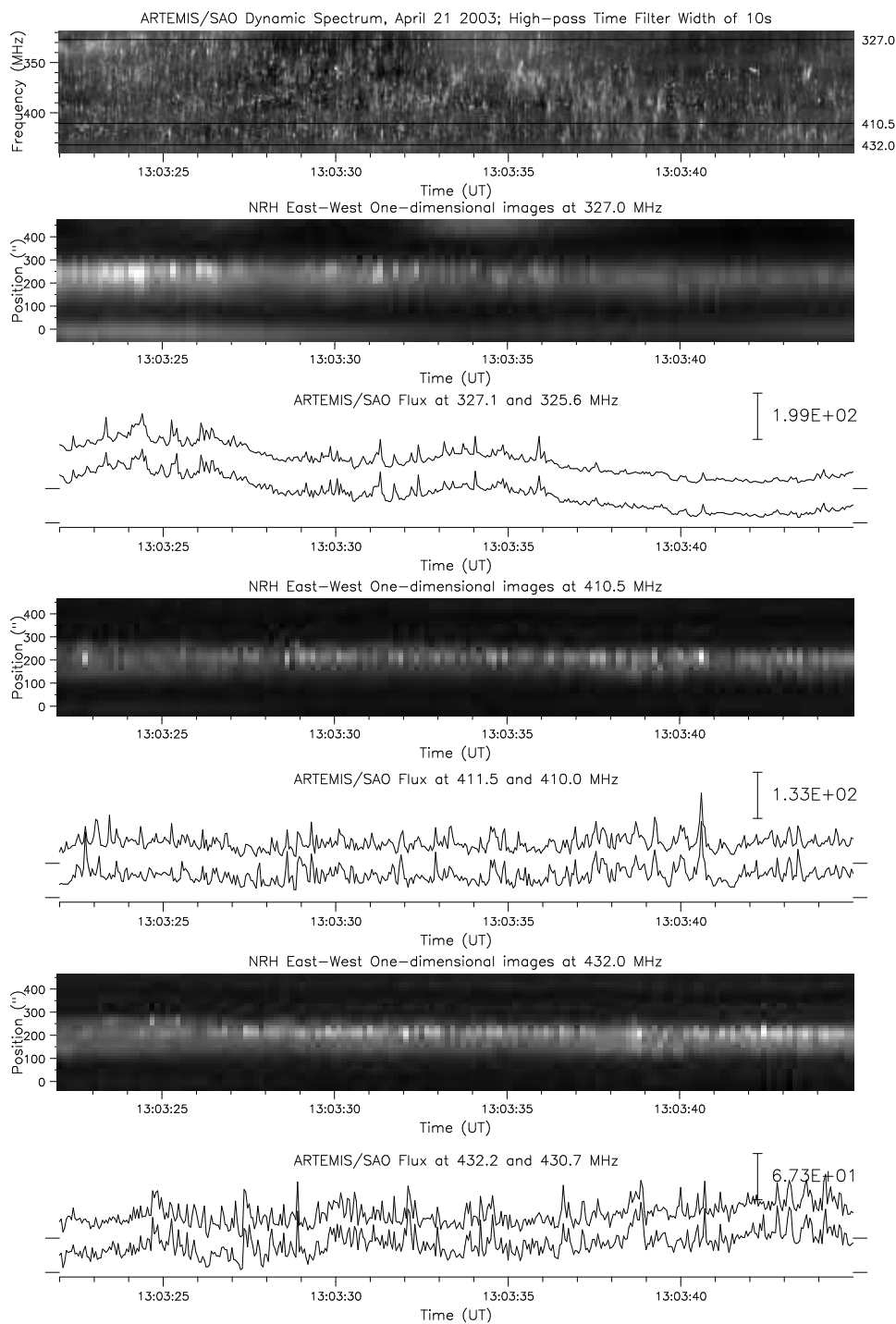}
\caption{Top rows: SAO dynamic spectrum. The position of the frequencies of the NRH are marked on the right. EW one-dimensional NRH images at 327\,MHz as a function of time. SAO time profiles at 327.1 and 325.6\,MHz; {\bf the bar gives the intensity scale in arbitrary units}. Other rows: same for 410.5 and 432\,MHz}
\label{saonrh1d}
\end{figure*}

\begin{figure*}
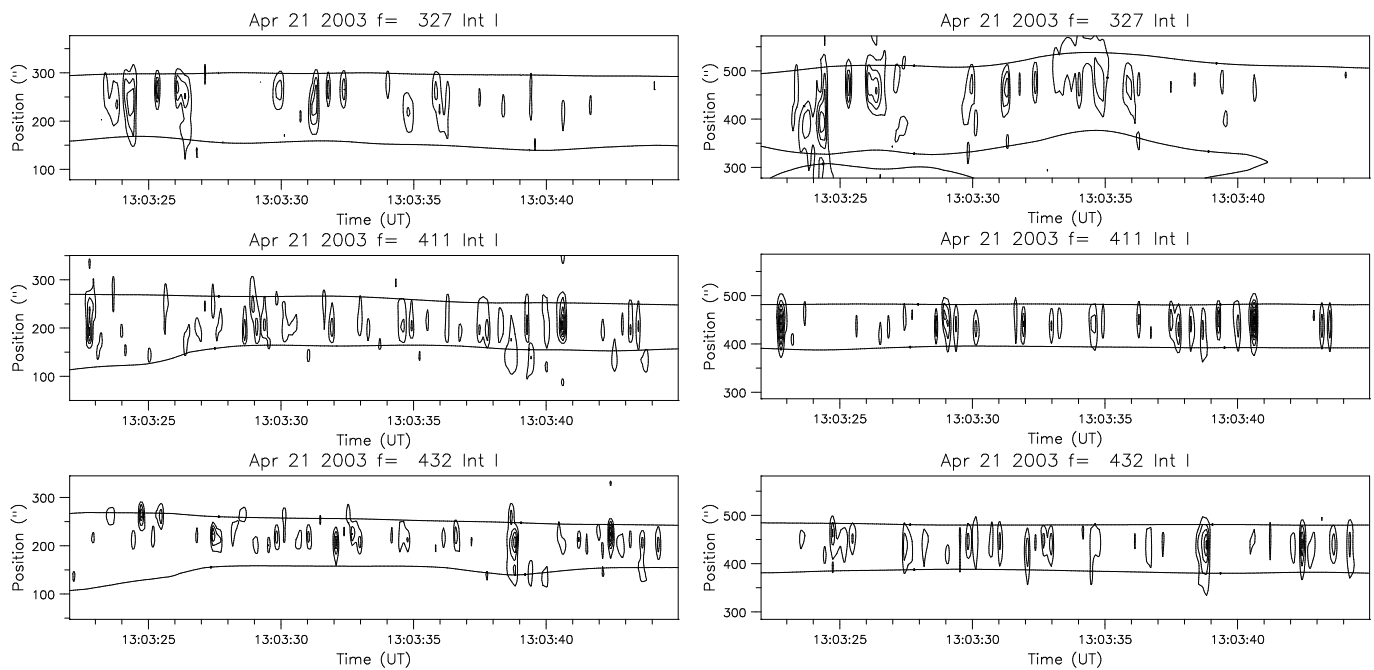

\centering
\includegraphics[angle=-90,width=.48\textwidth]{SPK_327EWC2.eps}~~~%
\includegraphics[angle=-90,width=.48\textwidth]{SPK_327NSC2.eps}
\includegraphics[angle=-90,width=.48\textwidth]{SPK_411EWC2.eps}~~~%
\includegraphics[angle=-90,width=.48\textwidth]{SPK_411NSC2.eps}
\includegraphics[angle=-90,width=.48\textwidth]{SPK_432EWC2.eps}~~~%
\includegraphics[angle=-90,width=.48\textwidth]{SPK_432NSC2.eps}
\caption{Contour plots of 1D intensity vs time after a  high-pass filtering in time, from the EW (left column) and the NS (right column) NRH arrays. The two lines along the time axis indicate the width of the background component.}
\label{filt}
\end{figure*}

A six-second long portion of the spike dynamic spectrum is presented in Fig. \ref{spikes2} at the full 10\,ms SAO time resolution. Although some type-III and U activity was present below $\sim300$\,MHz, it had no obvious association with the spikes, which occurred in a limited frequency range between 320 and 440\,MHz. The 2D autocorrelation function of the spectrum (shown as an insert in Fig. \ref{spikes2}) gave an average burst duration of 48\,ms and an average bandwidth of 7.7\,MHz, corresponding to a relative bandwidth of 2.1\% (Table \ref{TableSpikes}). These values fall near the peak of the 2D histogram (Fig. \ref{SpikeDuration}). 

Many spikes resemble tiny type-III bursts, and the autocorrelation image indicates an average frequency drift of -1800 MHz/s; this corresponds to a logarithmic drift of -4.9\,s$^{-1}$, which is higher than the values given in Sect. \ref{DurBW}. This value exceeds the upper limit of detectability estimated in Sect. \ref{FrequencyDriftRate} by $\sim20$\%. }
If this drift is interpreted in terms of exciter motion, it would require a local scale height less than 37.5\,Mm to keep the velocity of the exciter lower than the speed of light. As noted in Sect. \ref{FrequencyDriftRate}, density scales considerably smaller than the hydrostatic scale height of the quiet corona are not unlikely in the region of the type-IV emission. The light transit-time effect mentioned in the same section may have also affected the measured drift of this event, since it was located near the center of the solar disk. 

Several spikes occurred at the NRH frequencies that were within the SAO spectral range (327, 410.5, and 432\,MHz; \mbox{Fig. \ref{spikes2})}.  To compare the NRH data with the dynamic spectrum, we computed 1D images of intensity as a function of time using only visibilities from the EW and NS arrays. In addition to facilitating the comparison, this has the advantage of using the full resolution of the two arrays, whereas instantaneous 2D maps use only the inner part of the u-v plane
effectively ({\it cf.} Sect. \ref{Inst}). We furthermore note that, at the time of our observations, the orientation of the 1D EW images was 21.5\degr\ N of W and that of the NS images 11.5\degr\ E of N. 

The results for the EW array are presented in \mbox{Fig. \ref{saonrh1d}} for the time interval 13:03:22 to 13:03:45~UT, together with the SAO dynamic spectrum and time profiles of the SAO channels nearest to the NRH frequencies. The SAO data were integrated over 0.1\,s, in order to match the NRH time resolution of 0.15\,s. They were also subjected to a high pass filter in the time domain of 10\,s width.

In spite of the 150\,ms time integration of the NRH images, which is three times longer than the average duration of the spikes in this group, many spikes persist  in the SAO time profiles, and practically all of them are detectable in the NRH 1D images. At 327\,MHz, the spikes originate in the main source seen in Fig. \ref{sxinrh}, while the other two sources, visible in the upper and lower parts of the intensity-time plot of  \mbox{Fig. \ref{saonrh1d}}, show time variations on a longer time scale without any spikes. The same source was the origin of spikes at 410.5 and 432\,MHz.

\subsection{Position and size of spikes}
It is obvious from \mbox{Fig. \ref{saonrh1d}} that the spikes are superposed on a slowly varying background at all frequencies. To separate the spikes from the background, we made use of temporal filtering. To extract the background, we used a Gaussian low pass filter in the time domain with a  FWHM of 3\,s, whereas for spikes we employed a high pass filter of 1\,s width. In \mbox{Fig. \ref{filt}} we give contour plots of the high pass filtered 1D intensity as a function of time from the EW and the NS arrays. The figure shows that most spikes are smaller then the background source, and many of them are displaced with respect to it, always remaining within its half width \mbox{({\it cf.\/} Fig. \ref{saonrh1d}}}.

To quantify these properties we computed difference 2D images between some prominent spikes and the nearest background image and performed a Gaussian fit to the spike sources. The results are given in Table \ref{2003b}, where
the brightness temperature of the spikes was 1.2 to $1.9\times10^8$\,K, i.e. 0.5 to $3.5\times10^8$\,K above that of the background source, with the average temperature ratio around 0.7.

The area of spike sources is about 80\% that of the background, and on the average, they are shifted by 10 to 30\arcsec\  with respect to that. Although this shift is smaller than the NRH resolution, its reality is confirmed by inspecting the 1D images of Figs.  \ref{saonrh1d} and \ref{filt}, which show that spikes are often at the flanks of the slowly varying source.

\begin{table}
\begin{center}
\caption{Average parameters of spikes for the 2003 April 21 event}\label{2003b}
\begin{tabular}{lrrr}
\hline
Parameter                                               & 327   & 411   & 432     \\
\hline
Number of spikes                                        & 8             & 15      & 13    \\
Shift, \arcsec\                                 & 35    & 11    & 18    \\
Spike $T_b$, 10$^8$\,K                  & 1.2   & 1.9   & 1.6   \\
$T_b$ ratio, spikes/background  & 0.57  & 0.85  & 0.66  \\
Area ratio from 2D                              & 0.86  & 0.81  & 0.83  \\
Area ratio from 1D                              & 0.32  & 0.50  & 0.41  \\
Size from 1D, \arcsec\                          & 86    & 82    & 72    \\

\hline
\end{tabular}
\end{center}
\end{table}

More exact estimates of positions and sizes were obtained from the 1D images. Using the high and low pass images, computed as described above, as representative of the spikes and the background respectively, we found lower area ratios between 0.3 and 0.5. This is apparently due to the better resolution of the 1D images and shows that spikes are not fully resolved in the 2D images, having a characteristic size around 80\arcsec. Consequently, the observed brightness quoted above is a lower limit. The time integration of the NRH is an additional factor of the observed brightness being less than the true. We also note that, due to scattering effects, the true size is expected to be smaller than observed \citep[cf.][]{Mercier2015}. The results for each frequency are given in the last two rows of Table \ref{2003b}. %

The high value of brightness temperature of $10^{15}$\,K by  \citet{Benz1986} was based, however, on an estimated
size of only 200\,km (0.3\arcsec); regardless of that, our values of  $\sim2\times10^8$\,K exclude thermal emission and point toward 
non-thermal or coherent mechanisms.

The above analysis indicates that spikes are not caused by a fluctuation in the background source, but represent additional emission, such as what would be expected from small-scale reconnection within or very close to the background source.

\subsection{Polarization}
\begin{figure}[h]
\centering
\includegraphics[angle=-90,width=\hsize]{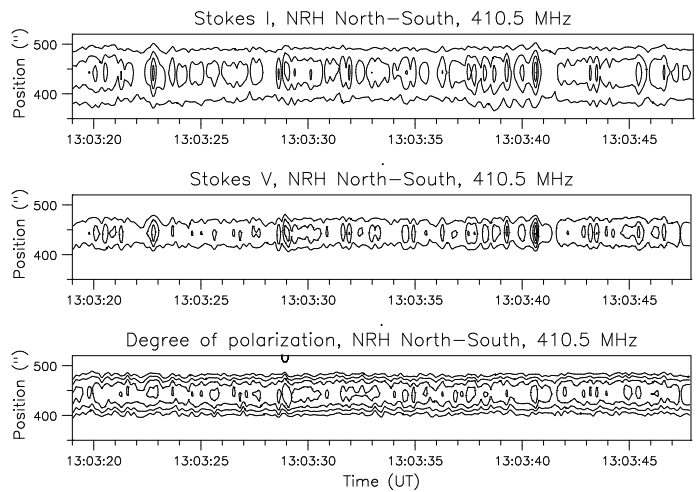}
\caption{Contour plots of 1D intensity as a function of position and time in Stokes I and V, as well as the degree of polarization. The polarization is in the left circular direction. Contours are in steps of 20\% of the maximum value.}
\label{IVP}
\end{figure}

The polarization calibration of the NRH is not very accurate and in our case it was rather bad. Still we managed to improve the quality through self-calibration and obtain meaningful results. An example is given in Fig. \ref{IVP}, where contours of 1D images in $I$, $V$ and the degree of polarization are plotted for the NS array at 410.5\,MHz. 

We first note that the spikes were polarized in the left circular direction, as was the background source. As reported in the literature \citep[e.g.,][]{Benz1986,Chernov2011}, the polarization of spikes is high, up to almost 100\%. Here the maximum is 66\%, but we note that this is comparable to the value of the background (50\%); in fact, spikes are much more prominent in the I and V images than in the polarization degree contours  (Fig. \ref{IVP}). Similar values were measured at 326.5 and 412\,MHz; in contrast, the polarization of the emission was very low, on the order of 10\% or less, at lower frequencies (150.9, 164.0, and 236.6\,MHz), where spikes did appear.

It is also obvious from Fig. \ref{IVP} that both spikes and the continuum have a smaller size in V than in I. We found that, on the average, the size of polarized sources was 60\% of that of total intensity.

\section{Summary and conclusions} \label{discussion}

High resolution dynamic spectra permit the recording and study of the hyperfine structure of radio bursts. So far only a  few cases of hyperfine structure have been reported  \citep{Kuznetsov2007, Chernov2012, 2012_TanBaolin&al}. Using the SAO receiver, operating in the frequency range of 450--270 MHz on the ARTEMIS-IV radio-spectrograph with a 10\,ms time resolution, we observed a large number of narrow-band bursts embedded in Metric Type--IV radio continua.
The high time resolution of the SAO receiver made the morphological distinction possible between a large number of different narrow-band structures, such as narrow-band J or U bursts and the unusual inverted U burst. 

As regards individual spike characteristics, such as bandwidth and duration, our measurements of more than 10\,000 spikes gave average values of 2\% and 100\,ms, respectively, which is consistent with past reports \citep{Csillaghy1993, Bouratzis10,  Dabrowski2011}. These values obey the empirical duration-frequency power laws proposed by \citet{Guedel1990}, \citet{Meszarosova03} and \citet{Rozhansky2008} and the instantaneous bandwidth-frequency power law of \citet{Csillaghy1993}. Although the duration-bandwidth histogram did not show any clear separation in subgroups, a small number (less than 1\%) of the sample consisted of broadband spikes, similar to those found by \citet{Bakunin85}. We also found a tendency of the bandwidth to increase with duration.

An examination of frequency drift rates for $\sim300$ spikes gave values of $-2.5\lesssim d\ln f/dt \lesssim4.5$\,s$^{-1}$. Negative drifts were found in 60\% of the cases with an average  $-1.12$\,s$^{-1}$, which is not much above type-III drift rates but about four times higher than spike-associated type-III bursts. 
Positive frequency drifts with an average value of 1.43\,s$^{-1}$  were measured for 21\% of the spikes, while the rest showed no measurable drift. Broadband spikes had drift rates that were an order of magnitude larger than the typical values. Spikes with high drift rates require density scale heights that are smaller than predicted from quiet-Sun models, as expected in local condensations, in order to keep the exciter velocity lower than the speed of light.

Quite often spikes are not scattered in the dynamic spectra, but align in groups. Column spikes cluster within a short time interval of $\sim1$\,s over a wide frequency range and have positive or negative drift rates of 1.36 to $-3.25$\,s$^{-1}$, which is comparable to the standard type-III bursts. Chains, on the other hand, have a group frequency drift comparable to the intermediate drift bursts (fibers) in the range of 0.033 to $-0.021$\,s$^{-1}$. In addition to columns and chains, we detected bi-directional spikes in the form of two parallel chains, as well as spike groups that were aligned in the form of N bursts and Lace bursts. Obviously, such groups would have appeared as continuous structures under lower time resolution.

Finally, we were able to image spikes with the Nan\c cay Radioheliograph at three frequencies within the SAO range. The high time and angular resolution of the NRH made the detailed investigation of the brightness of spikes, their position. and size possible. The spikes appeared on top of a slowly varying background source, located 100 to 240\arcsec\ NW of the soft X-ray flare source imaged by GOES/SXI. Two weaker nearby sources had no spike-associated emission. We measured brightness temperatures of 1 to 2$\times10^8$\,K above the background. The spikes were smaller and were often displaced with respect to the background, indicating that they are not fluctuations of the background intensity, but represent additional emission such as what would be expected from small-scale reconnection within or very close to the background source. Both spikes and the background source were highly polarized in the same sense at about 65\%, while the average size of polarized emission was $\sim40$\% smaller than that of the total intensity.

\begin{acknowledgements}
This work was supported in part by the Special Account for Research Grants of the National and Kapodistrian University of Athens. The NRH (Nan\c cay Radioheliograph) is operated by the Observatoire de Paris and funded by the French research agency CNRS/INSU. The authors are grateful to the colleagues at Meudon for providing the original NRH visibilities. 
\end{acknowledgements}
\bibliographystyle{aa}

\end{document}